\documentclass{aa}
\usepackage{graphicx}
\usepackage{txfonts}
\usepackage{hyperref}
\hypersetup{colorlinks = true, citecolor = {blue}, urlcolor= {blue}}

\def\PGPU{$\varphi-$GPU}

\def\gapprox{\;\rlap{\lower 3.0pt                       
        \hbox{$\sim$}}\raise 2.5pt\hbox{$>$}\;}
\def\lapprox{\;\rlap{\lower 3.1pt                       
        \hbox{$\sim$}}\raise 2.7pt\hbox{$<$}\;}

\newcommand{\be}{ \begin{equation} }
\newcommand{\ee}{\end{equation}}

\newcommand{\ben}{\begin{enumerate}}
\newcommand{\een}{\end{enumerate}}

\renewenvironment{thebibliography}[1]{\begin{oldthebibliography}{#1}\setlength{\parskip}{0ex}\setlength{\itemsep}{0ex}}{\end{oldthebibliography}}

\usepackage{orcidlink}
\let\orcid\orcidlink
\makeatletter
\renewcommand*\aa@pageof{, page \thepage{} of \pageref*{LastPage}}
\makeatother

\def\ba{\mbox{\boldmath $a$}}
\def\acck#1{\ba_i^{(#1)}}
\def\absak#1{|\acck{#1}|}

\begin{document}

\title{Cosmological insights into the early accretion of r-process-enhanced stars \\ II. Dynamical identification of lost members of Reticulum II}

\author{Peter~Berczik
\inst{1,2,3}\orcid{0000-0003-4176-152X}
\and
Maryna~Ishchenko
\inst{1}\orcid{0000-0002-6961-8170}
\and
Olexandr Sobodar
\inst{1,2}\orcid{0000-0001-5788-9996}
\and
Mohammad Mardini
\inst{4,5}\orcid{0000-0001-9178-3992}
}

\institute{Main Astronomical Observatory, National 
Academy of Sciences of Ukraine,
27 Akademika Zabolotnoho St, 03143 Kyiv, Ukraine  \email{\href{mailto:berczik@mao.kiev.ua}{berczik@mao.kiev.ua}}
\and
Nicolaus Copernicus Astronomical Centre Polish Academy of Sciences, ul. Bartycka 18, 00-716 Warsaw, Poland
\and
Konkoly Observatory, Research Centre for Astronomy and Earth Sciences, HUN-REN CSFK, MTA Centre of Excellence, Konkoly Thege Mikl\'os \'ut 15-17, 1121 Budapest, Hungary
\and
Department of Physics, Zarqa University, Zarqa 13110, Jordan
\and
Jordanian Astronomical Virtual Observatory, Zarqa University, Zarqa 13110, Jordan
}
  
\date{Received xxx / Accepted xxx}

\abstract
{}
{We identify the possible dynamical connection between individual \textit{r}-process-enhanced stars and the ultra-faint dwarf galaxy Reticulum II based on the current phase-space information for these stars and the dynamical mass-loss model of Reticulum II during its orbital motion for 11.5\, Gyr of lookback time.
The dynamical orbital modelling together with the chemical abundance analysis proved to be useful tools for the progenitor identification of the peculiar stars  in our Galaxy.}
{To reproduce the Reticulum II orbital mass loss, we used our high-precision $N$-body \PGPU\ code to integrate almost 1 million stars into the system evolution inside a external Galactic potential. We also investigated the orbits of \textit{r}-process-enhanced stars using the same code.} 
{We present our Reticulum II dynamical modelling results in the context of the stars' energies -- angular momentum phase-space and phase-space overlapping of the currently observed \textit{r}-process-enhanced stars with Reticulum II stellar tidal tails. Of the 530 \textit{r} stars known today, at least 93 are former members of the Reticulum II dynamical progenitor system.}
{}


\titlerunning{Ret-II with \textit{r}-stars}
\authorrunning{P.~Berczik et al.}
\maketitle

\section{Introduction}\label{sec:Intr}

Dwarf galaxies serve as cosmic time capsules, revealing secrets about the Universe's early days and the formation of larger galaxies, such as the Milky Way \citep[e.g.][]{Frebel_2010_nat}. Their interactions and mergers help us understand how our Galaxy grows and evolves over time \citep[e.g.][]{Martin_2007, Geha2009,Chiti2024NatAs}. They also offer a unique window into star formation and the spread of elements throughout the cosmos \citep[e.g.][]{Chiti2023AJ,Xiaowei2024}. Dwarf galaxies appear to have formed all their stars in their first $\sim$1\, Gyr, before star formation is cut off by reionisation \citep{Benson2002, Brown2014, Ji2023}. In particular, ultra-faint dwarf  (UFD) galaxies are at the extreme low-mass end of galaxy formation, where star formation is inefficient and massive stars form sporadically; this results in intermittent feedback and incomplete sampling of nucleosynthetic sources \citep{Koch2013, Frebel2010, Frebel2015, Ji2019}. They are key pieces of the puzzle and help us unravel the story of our Universe, providing a unique window into the first stars and galaxies in the pre-reionisation Universe \citep{Mayer2023, Mayer2024}.

One example is Reticulum II (hereafter Ret-II), discovered by the Dark Energy Survey \citep{Bechtol2015, Koposov2015}.\ It is located at a distance of approximately 30--32 kpc from the Galactic Center and has an age of around 11.5--12.5\, Gyr, which is typical for a UFD galaxy \citep{Simon2023}. Ret-II is one of the prototypical UFD satellites of the Milky Way. It has a metallicity range of \mbox{$-$3.5 $<$ [Fe/H] $< -2 $}. Interestingly, about $\sim$70\% of Ret-II stars exhibit an enhanced $r$-process (\textit{r}-PE) signature  \citep{Roederer2016, Frebel2019, Simon2023}, which suggests that it may have experienced a unique nucleosynthetic event, such as a neutron star merger or a rare supernova. Therefore, it is crucial to understand this UFD's star formation conditions to interpret the properties of the most metal-poor stars.

In addition, exploring the motion and morphology of Ret-II will help us understand the complex formation and evolution history of our Galaxy \citep{Brauer2021,Brauer2022}. Therefore, linking the observed chemical signatures in Ret-II with orbital histories of the halo \textit{r}-PE can be a powerful tool for placing crucial constraints on the merger tree (hierarchical structure) of the Milky Way. Various efforts have been made to build comprehensive astronomical catalogs, such as the ESA/\textit{Gaia} astrometric mission \citep{Gaia_the_mission}. The 6D astrometric data provided in \textit{Gaia} Data Release 3 \citep[DR3;][]{Gaia_DR3} allow us to accurately reconstruct the orbital history of the known $r$-PE stars in our Galaxy over the past billion years.

Our main goal in this work is to identify the possible dynamical connection between observed \textit{r}-PE stars and the Ret-II UFD galaxy based on the current phase-space information for these stars and the dynamical mass-loss model of the Ret-II object during its orbital motion relative to our Galaxy (11.5\, Gyr of lookback time). To reproduce the Ret-II orbital mass loss, we used our high-precision $N$-body \PGPU (graphics processing unit) code to integrate the dynamical evolution of the Ret-II system with almost 1 million stars inside the external Galactic potential. We also integrated the orbits of observed \textit{r}-PE stars using the same code inside the same external Galactic potential. To get a physical understanding, we placed Ret-II and the selected stars in a time-varying Milky Way-like potential where the mass and scale parameters of the Galactic external potential dynamically change over time \citep{Ishchenko2023a}. 

Our paper organised as follow. In Sect. \ref{sec:init-integr} we describe the time-variable external potential, initial conditions, and orbital reconstruction of Ret-II and the \textit{r}-PE stars. In Sect. \ref{sec:ret-no-sse} we show the star-by-star dynamical orbital evolution for Ret-II over the whole 11.5 billion years. In Sect. \ref{sec:phase-space} we present our Ret-II dynamical modelling results in the context of the stars' energy -- the angular momentum phase-space and phase-space overlapping of the known \textit{r}-PE stars with the Ret-II stellar tidal tails. 

\section{Initial data and integration procedure}\label{sec:init-integr}

\subsection{Time variable potential}\label{subsec:orb-ret}

\begin{figure*}[htbp]
\centering
\includegraphics[width=0.95\linewidth]{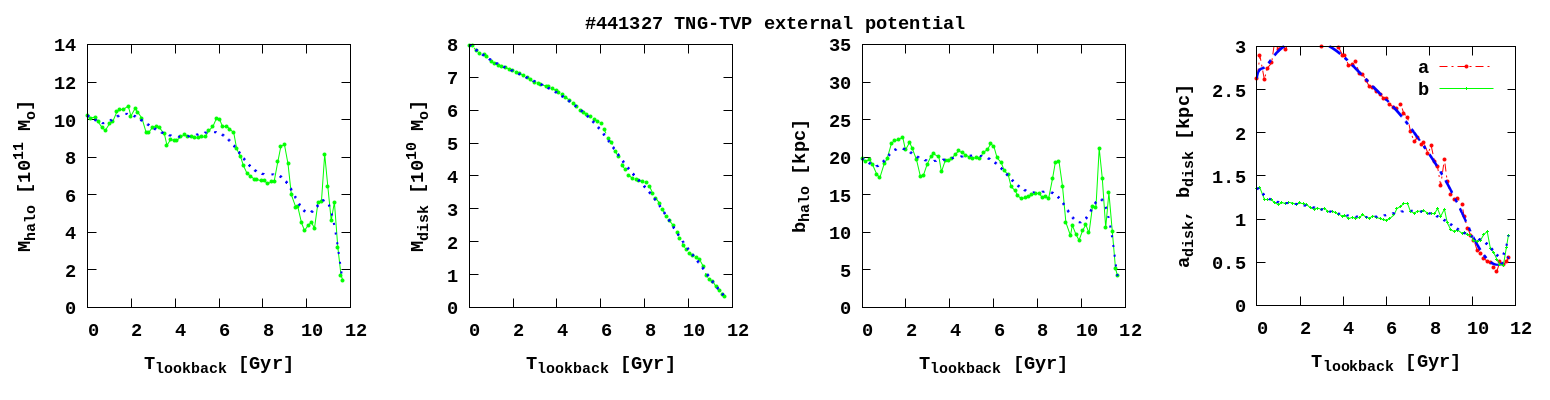}
\caption{Evolution of halo and disk masses, and their characteristic scales, for TNG-TVP {\tt 441327}. The halo mass ($M_{\rm h}$), disk mass ($M_{\rm d}$), halo-scale parameter ($b_{\rm h}$), and disk-scale parameters ($a_{\rm d}$ and $b_{\rm d}$) are presented \textit{from left to right}. Solid green lines with dots show the parameters derived from the IllustrisTNG-100 data. Dotted and dash-dotted blue lines correspond to the values after interpolation and smoothing with a 1~Myr timestep.}
\label{fig:MW-TNG}
\end{figure*}

We decided to model the Ret-II dynamical system starting from 11.5\, Gyr of lookback time based on the current age estimation for this system \citep{Simon2023}. For such a long dynamical modelling of the orbital evolution of Ret-II in our Galactic field, we clearly need a realistic description of the Galactic potential. For this reason, we used our standard $N$-body code and the time-dependent external potential described in \citet{Mardini2020}. This potential was derived from the IllustrisTNG-100 cosmological simulation database \citep{Nelson2018, Nelson2019, NelsonPill2019}. The procedures for sampling and fitting these potentials are described in detail in \citet{Mardini2020, Mardini2022_j1808} and \citet{Ishchenko2023a}. The numerical code and accompanying routines are also publicly available on GitHub.\footnote{The ORIENT:\\~\url{https://github.com/Mohammad-Mardini/The-ORIENT}} For our task we selected the {\tt 441327} time variable potential (TNG-TVP), which (at redshift $z = 0$) has parameters quite similar to those of our present-day Galaxy, for example halo and disc masses and their characteristic scales (see Table~\ref{tab:pot} and Fig. \ref{fig:MW-TNG}). 

\begin{table}[htbp!]
\caption{Parameters of the TNG-TVP at redshift zero.}
\centering
\resizebox{0.49\textwidth}{!}{
\begin{tabular}{llccc}
\hline
\hline
\multicolumn{1}{c}{Parameter} & Unit & {\tt 441327} &  Milky Way \\
\hline
\hline
Disc mass,  $M_{\rm d}$         & $10^{10}~\rm M_{\odot}$ & 7.97 &  6.79 \\
Halo mass,  $M_{\rm h}$         & $10^{12}~\rm M_{\odot}$ & 1.02 &  1.00 \\

Disc scale length, $a_{\rm d}$ & 1~kpc                     & 2.63 & 3.41 \\
Disc scale height, $b_{\rm d}$ & 1~kpc                     & 1.36 & 0.32 \\
Halo scale height, $b_{\rm h}$ & 10~kpc                    & 1.98 & 2.77 \\
\hline
\end{tabular}
}
\tablefoot{The last column shows the parameters of the corresponding Milky Way components according to \cite{Bennett2022} and \cite{Bland-Hawthorn2016}.}
\label{tab:pot}
\end{table} 

\subsection{Orbital reconstruction in ORIENT for  Ret-II}\label{subsec:orb-ret1}

We used the positions and velocities for Ret-II in the equatorial coordinate system at present day from \cite{Koposov2015} and \citet[][see our Table \ref{tab:init-data}]{Walker2015}. To find the dynamical integration of the Ret-II orbit using our $N$-body code, we needed to convert these quantities to the Cartesian galactocentric frame. We assumed the distance to the Sun from the Galactic Centre at the plane to be $X_{\odot} = -8.178$~kpc \citep{Gravity2019} with an above-plane height of $Z_{\odot} = 20.8$~pc \citep{Bennett2019}. The velocity transformation is described in \cite{Johnson1987}. For the equatorial position of the north galactic pole (NGP), we used the updated values from \cite{Karim2017}: RA$_{\rm NGP} = 192\fdg7278$, Dec$_{\rm NGP} =  26\fdg8630$, and $\theta_{\rm NGP}  = 122\fdg9280$. For the local standard of rest~(LSR) velocity, we used 234.737~km~s$^{-1}$ \citep{Reid2004}, and for the peculiar velocity of the Sun with respect to the LSR, we used $U_{\odot} = 11.1$~km~s$^{-1}$, $V_{\odot} = 12.24$~km~s$^{-1}$, and $W_{\odot} = 7.25$~km~s$^{-1}$~\citep{Schonrich2010}. Since the integration was performed backwards in time, the sign of the velocity components for Ret-II was changed to the opposite value:

$X$ =  $-$9.4, kpc \;\;\;\;\;\;\;\;\;\; $V_x$ = $-$30.9, km s$^{-1}$ \;\;\;\;\;\;\;\;\;\;\;\;\;\;\;\;\;\;\;\;\;\;\;\;\;

$Y$ = $-$19.3, kpc \;\;\;\;\;\;\;\; $V_y$ = 85.9, km s$^{-1}$ \;\;\;\;\;\;\;\;\;\;\;\;\;\;\;\;\;\;\;\;\;\;\;\;\;\;\;\;\;

$Z$ = $-$22.8, kpc \;\;\;\;\;\;\;\; $V_z$ = $-$205, km s$^{-1}$. \;\;\;\;\;\;\;

\begin{table}[tbp]
\setlength{\tabcolsep}{4pt}
\centering
\caption{Ret-II positions in the equatorial coordinate system at the present day.}
\label{tab:init-data}
\begin{tabular}{lcc}
\hline
\hline 
 Parameter & Value & Unit \\
\hline
\hline
$\alpha$ & 53.9256 & deg \\ 
$\delta$ & -54.0492 & deg \\ 
$D_{\rm \odot}$ & 30.0 & kpc \\ 
RV & 62.8 & km s$^{-1}$ \\
$l$ & 266.2957 & \\
$b$ & -49.7357 & \\
$\mu_{\rm \alpha}$ & 2.34 & mas/yr \\
$\mu_{\rm \delta}$ & -1.31 & mas/yr \\
\hline 
\end{tabular}
\vspace{6pt}
\end{table}

For the orbital integration of Ret-II, we used our own high-order parallel $N$-body code \PGPU\footnote{$N$-body code \PGPU:\\~\url{ https://github.com/berczik/phi-GPU-mole}}, which is based on the fourth-order Hermite integration with a hierarchical individual block timestep scheme \citep{Berczik2011,BSW2013}. This code is well tested and has already been used to obtain important results in our previous globular cluster simulations \citep{Ishchenko2024mass-loss}. For the orbital reconstruction, we described Ret-II as a point mass with  $10^{6}$ $M_{\rm \odot}$ \citep{Walker2009, Wolf2010, Minor2019} and used the same TNG-TVP. 

\begin{figure*}[htbp]
\centering
\includegraphics[width=0.95\linewidth]{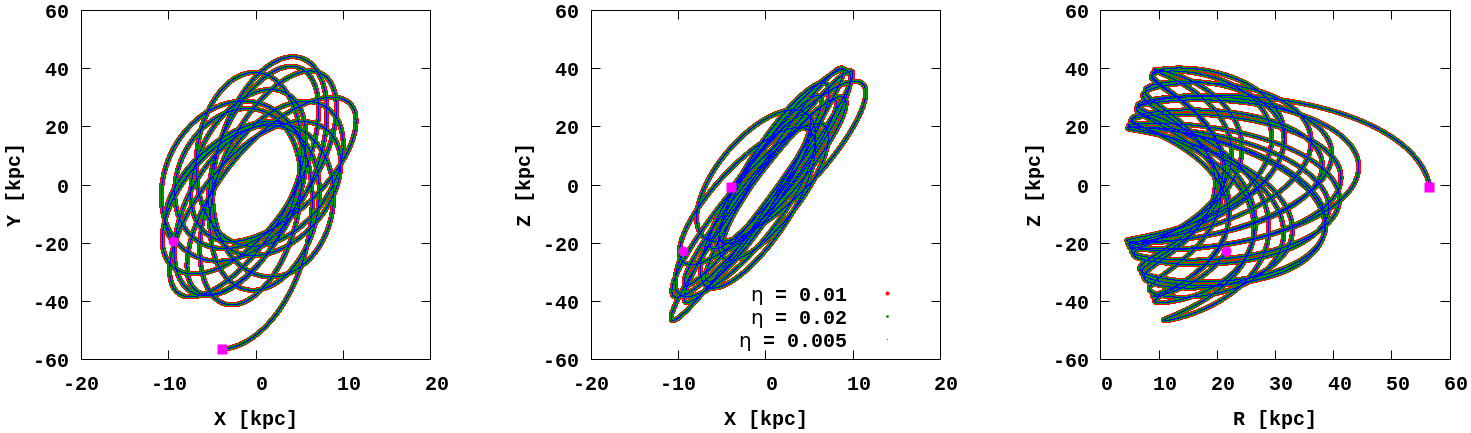}
\caption{Ret-II orbital reconstruction with 11.5\, Gyr of lookback time integration with different $\eta$ parameters. The magenta square shows the past position  and the circle the present-day position.}
\label{fig:dccoprr-back}
\end{figure*}

For an accurate orbital integration, the value of the timestep parameter, $\eta,$ can be quite  significant \citep{MA1992}. In the particular case of the fourth-order Hermite integration scheme, the timestep can be written as
\begin{eqnarray}
\label{eq:aarseth-timestep}
\Delta t_{i} &=& \sqrt{\eta} \cdot \frac{A_i^{(1)}}{A_i^{(2)}}, \\
A_i^{(k)} &=& \sqrt{\absak{k-1}\absak{k+1} + \absak{k}^2},
\end{eqnarray}
where $\boldsymbol{a}^{(k)}$ is the $k$-th derivative of the $i$-th particle acceleration. Thus, the timesteps are directly proportional to the $\eta$ parameter, which is responsible for the total integration accuracy. For the higher-order  Hermite integration schemes (sixth or eighth order), the generalised Aarseth criterion can be found in \cite{Nitadori2008}. We carried out the orbital integration for three different $\eta$ parameters: 0.01, 0.02, and 0.005. For all three runs, we get very consistent coordinates and velocity results for the 11.5\, Gyr lookback integration. The maximum discrepancy after this 11.5\, Gyr integration was around $\sim$0.1\%; this was for the $Z$ component of the coordinate and velocity. In Fig. \ref{fig:dccoprr-back} we present the Ret-II orbital reconstruction from 11.5\, Gyr of lookback time integration with different $\eta$ parameters. As we see, despite different $\eta$,  we have almost identical orbits. We used the $\eta$ = 0.01 value as the basic parameter for all the subsequent integrations.

\subsection{Initial conditions for \textit{r}-PE}\label{subsec:ini-r}

We selected our sample of metal-poor stars from JINAbase (a database for metal-poor stars) (\citealt{jinabase}\footnote{JINAbase: \\~\url{https://github.com/Mohammad-Mardini/JINAbase}}), a comprehensive database that compiles a wide array of atmospheric abundance measurements for metal-poor stars from various literature sources. We chose JINAbase over other available databases due to its extensive coverage, accuracy, and the detailed information it provides on a range of elements.

From JINAbase, we selected 1069 stars with available measurements for the elements europium (Eu), barium (Ba), and strontium (Sr), as these elements are crucial for identifying $r$-PE stars \citep[see ][for more details]{Mardini2024,Mardini2024ump}. After applying these criteria, we further refined our sample to include only stars classified as $r$-I (336 stars), $r$-II (146 stars), and $r$-limited ($r$-lim; 46 stars). These classifications are based on the relative abundance criteria described in \citet{Frebel2018ARNPS}. The groups $r$-I, $r$-II, and $r$-lim are identified as follows:

\textbf{$r$-I}: $0.3 \le \mbox{[Eu/Fe]} \le +1.0$ and $\mbox{[Ba/Eu]} < 0.0$; 

\textbf{$r$-II}: $\mbox{[Eu/Fe]} > +1.0$ and $\mbox{[Ba/Eu]} < 0.0$;

\textbf{$r$-lim:}  $\mbox{[Eu/Fe]} < 0.3$, $\mbox{[Sr/Ba]} > 0.5$, and  $\mbox{[Sr/Eu]} > 0.0$.

We then retrieved the astrometric solutions and their associated uncertainties from \textit{Gaia} DR3 \citep{Gaia_DR3} and corrected the parallaxes for the known bias as suggested by \citet{Lindegren_Parallax_2021}. We then calculated distances using a decreasing space density prior as described in \citet{Mardini2022Atari}. We also took the observational errors  into account when estimating the values of the relative errors for the components heliocentric distance (e$D_{\rm \odot}$), radial velocity (eRV) and proper motions (e$\mu_{\rm \alpha}$ with e$\mu_{\rm \delta}$). We finally investigated the possible influence of these relative errors on the integrated stellar orbits. 

\begin{figure*}[htbp]
\centering
\includegraphics[width=0.48\linewidth]{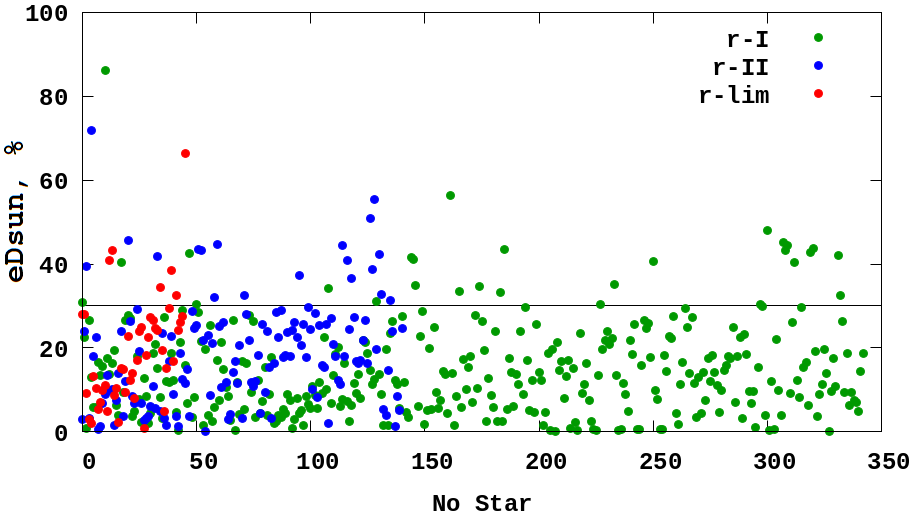}
\includegraphics[width=0.44\linewidth]{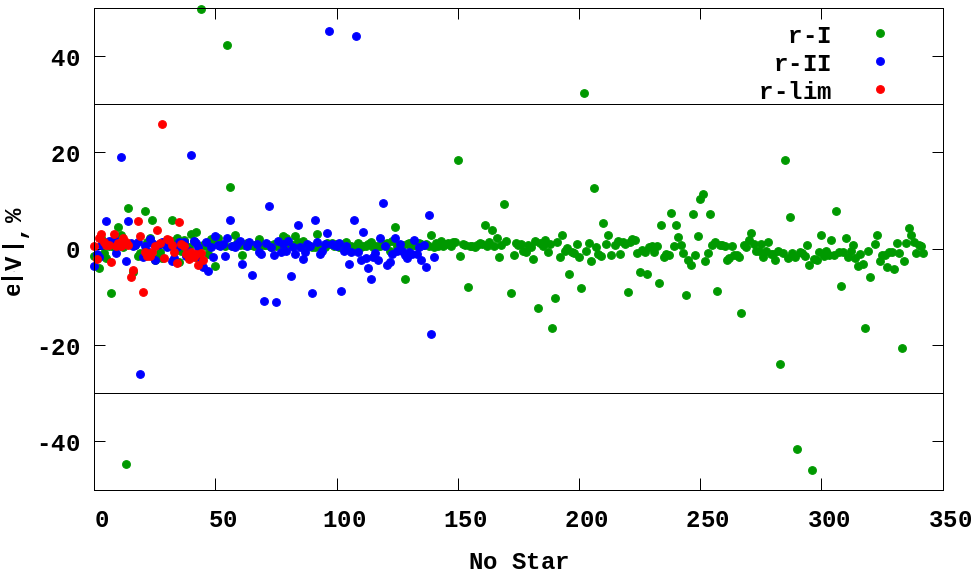}
\caption{Distribution of the relative errors in percent for the $D_{\rm \odot}$ and 3D velocity. Grey lines show the 30\% limiting values.}
\label{fig:star-error}
\end{figure*}

In Fig. \ref{fig:star-error} we present the relative error distribution in percent for $D_{\rm \odot}$ and relative to the 3D velocity in the galactocentric reference frame (i.e. including RV, $\mu_{\rm \alpha}$, and $\mu_{\rm \delta}$). Comparing the two panels, we see that the e$D_{\rm \odot}$ errors are distributed in a much wider range compared to the 3D velocity errors. At the same time, for both observations, we have few stars with relative errors greater than 100\%. 

\begin{figure*}[htbp]
\centering
\includegraphics[width=0.49\linewidth]{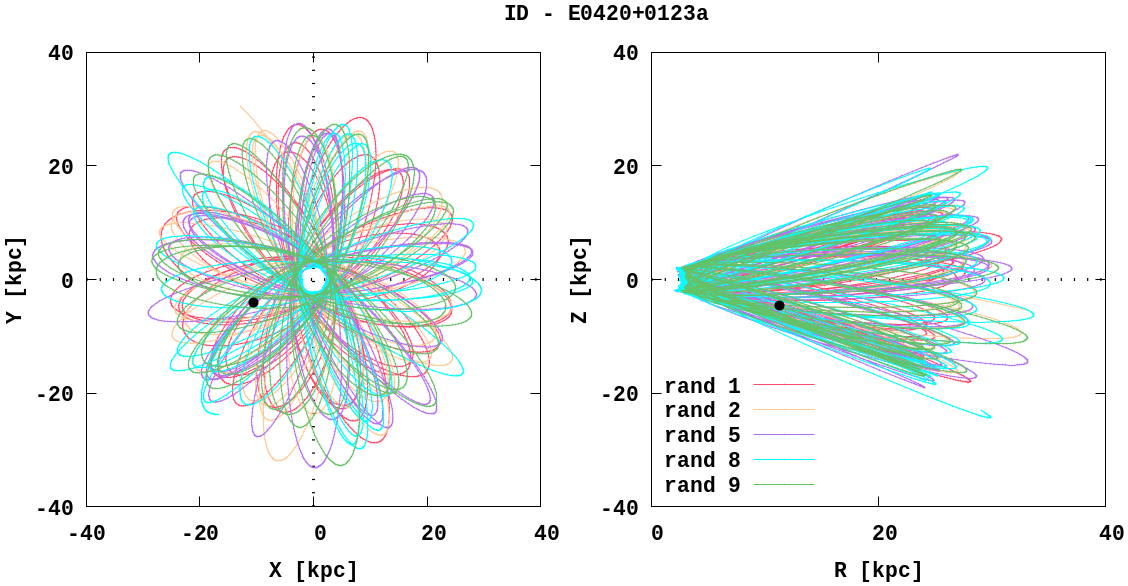}
\includegraphics[width=0.49\linewidth]{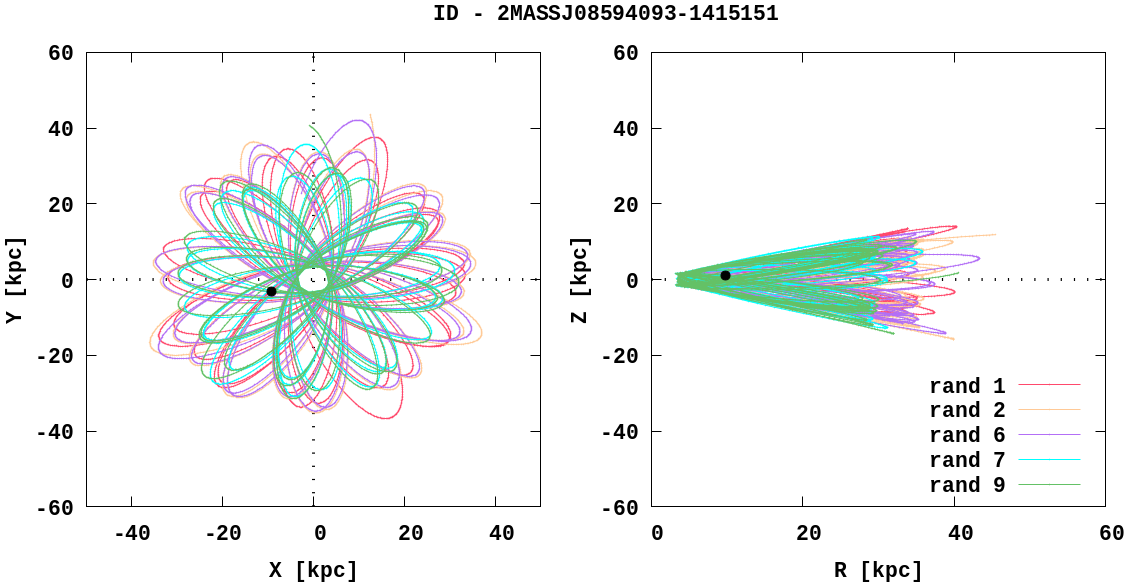}
\includegraphics[width=0.49\linewidth]{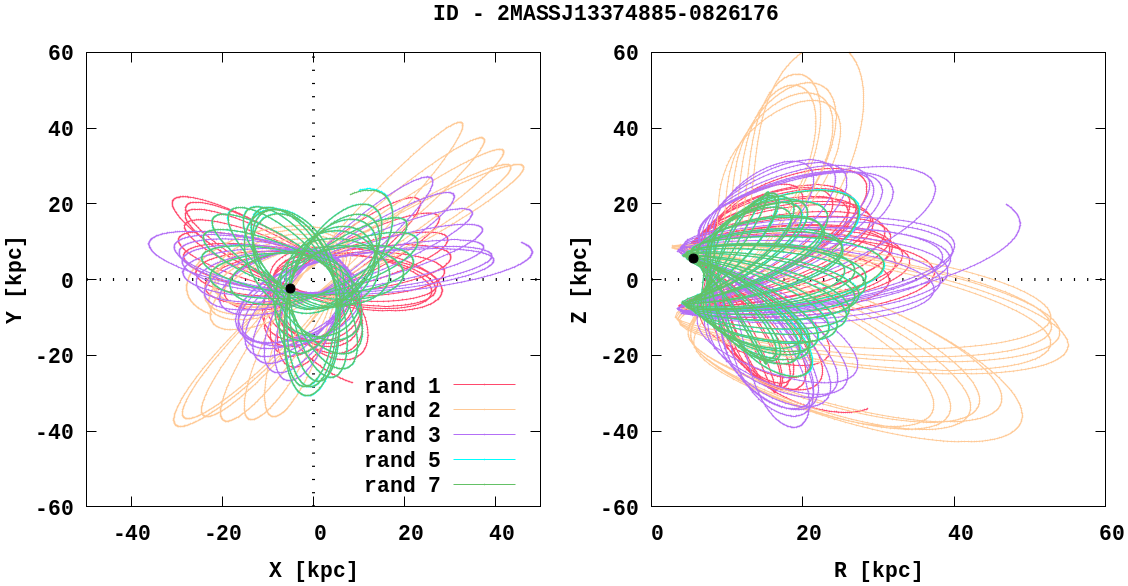}
\includegraphics[width=0.49\linewidth]{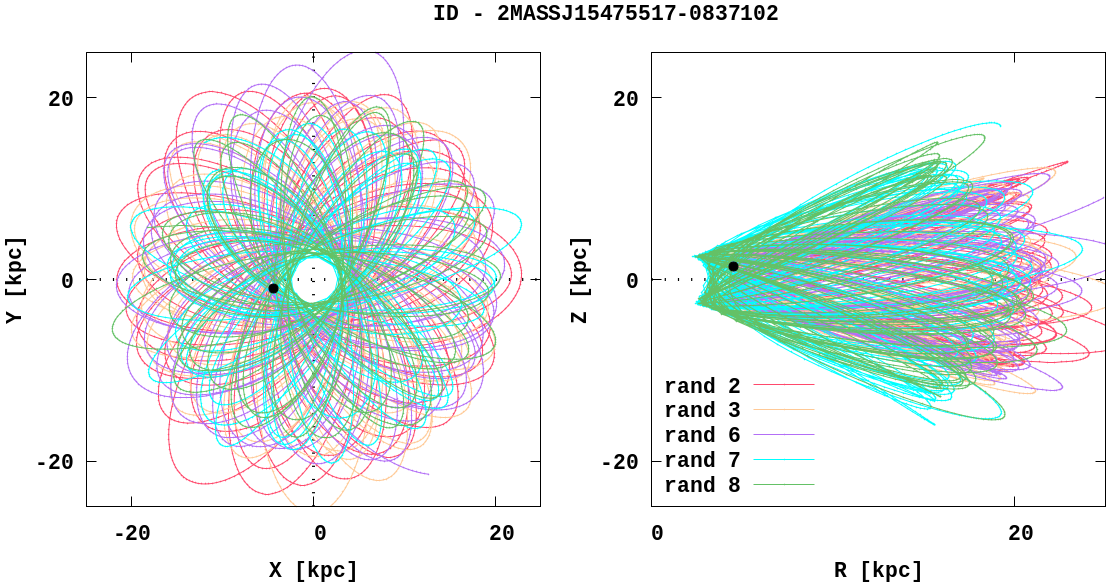}
\caption{Orbital reconstruction for four stars and five different random realisations (colour-coded) with 11.5\, Gyr of lookback integration evolution in {\tt 411321} TNG-TVP external potential.}
\label{fig:star-otb}
\end{figure*}

To check the influence of the observational errors on the stars' orbital parameters, we ran ten simulations, varying the initial distances and velocity components of the stars within their  $\pm \sigma$ error bar using the normal distribution. In Fig. \ref{fig:star-otb} we illustrate the impact of the initial condition uncertainties on the orbital evolution of four selected \textit{r}-PE stars for five random realisations in the {\tt 441327} TNG-TVP external potential. All the stars in our sample have a relative error of less than 30\%.    
 
\section{Ret-II star-by-star dynamical evolution}\label{sec:ret-no-sse}

As a basic physical model for Ret-II at 11.5\, Gyr ago, we chose the equilibrium King model \citep{King1962} with concentration parameter W0 = 8. In our numerical modelling we tried to reproduce an object that has nearly the same current physical parameters as Ret-II (i.e. similar masses and half-mass radii). As a basis for our physical parameters, we used the data presented in \cite{Simon2015, Simon2023}. Here we see a maximum mass estimation of $\sim$8 $\cdot$ 10$^5$ M$\odot$ and an estimated half light radius of around  $\sim$50 pc (which gives us a resulting 3D half-mass  radius of $\sim$35 pc). We chose a King concentration parameter of W0 = 8 to initially have a maximally compact King object in terms of the half-mass  radius to tidal radius ratio;  for this W0 value, the r$_{\rm hm}$/r$_{\rm tid}$ is close to the theoretical minimum of $\sim$0.1.

With such a parameter, the set of normalised King models has the minimum half-mass  to King radius ratio and consequently the smallest dynamical mass-loss  rate compared to the other concentration parameters. For the physical mass and radius normalisation, we chose a mass $M_{\rm ini} = 10^{6}$ and a half-mass radius of 44 pc. Based on our early TNG-TVP Galaxy model and the initial position of our UFD object, we estimate the proto-Ret-II Jacoby (tidal) radius \citep{Just2009, Ernst2011} to be $\approx$ 700 pc. The King radius (i.e. 100\% of the Lagrange radius of the model) for our model is $\approx$ 400 pc (i.e. our initial dynamical model is a significantly under-filled concentrated model).

For this physical model, we generated a set of numerical realisations, varying the initial total number of the particles over a wide range: $N = 10k, 20k, 40k, 80k, 160k, 320k, 640k$, and $960k$. We used this wide selection to find an appropriate range in which the dynamical mass loss from the object is basically saturating (i.e. where we no longer see the significant `N dependence' of the dynamical mass loss from the objects). The entire model integration was carried out from -11.5\, Gyr to today with the \PGPU $N$-body code, which is described in Sect. \ref{subsec:orb-ret}. 

Because our main interest in the modelling was a dynamical study of the orbits of escaped sub-solar-mass, long-lived stars in cosmologically motivated time and a variable external Galactic gravitational field (IllustrisTNG-100, {\tt 441327}), we neglected the stellar mass evolution of the generated $N$-body system. It is clear that for a deeper study of the internal evolution of Ret-II's stellar mass loss, the time evolution is also important. But in our case, where we are mainly interested in escaped stars with long lifetimes (of the order of 10\, Gyr) on independent Galactic orbits, we can neglect the stellar mass-loss time dependence. 

\begin{figure*}[htbp]
\centering
\includegraphics[width=0.32\linewidth]{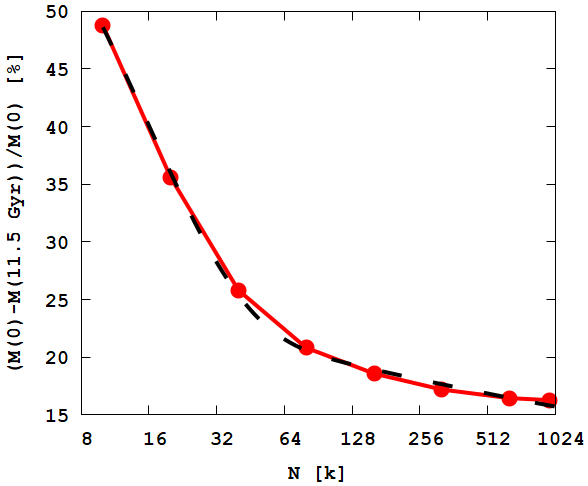}
\includegraphics[width=0.34\linewidth]{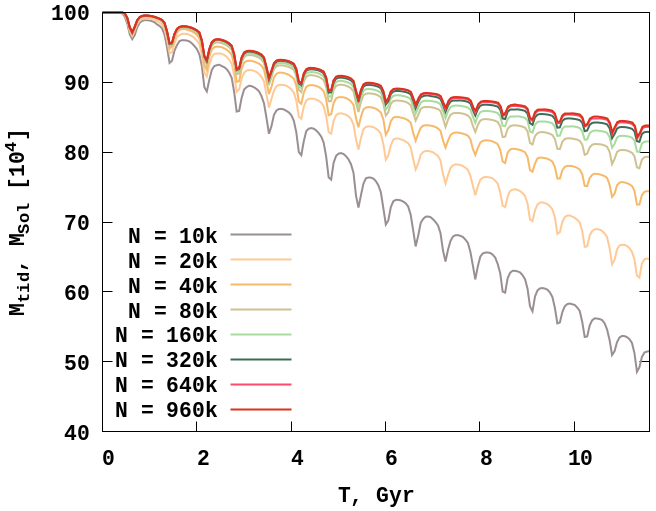}
\includegraphics[width=0.33\linewidth]{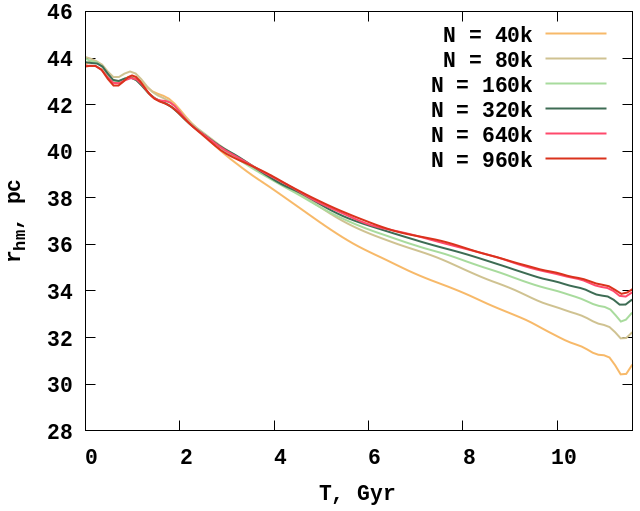}
\caption{Ret-II tidal mass and half-mass radius evolution for different models with different particle numbers. {\it Left panel}: Global mass loss (red) in percent for different particle numbers. The dashed black line represents the exponential fitting of the mass-loss  data for different $N$.}
\label{fig:mtid}
\end{figure*}

In Fig. \ref{fig:mtid} we present the evolution of the tidal mass for different numerical models. As we expect, for a large enough $N$, the models' mass losses are saturated (i.e. the number of escaped stars becomes almost $N$-independent and is on the level of $\approx$18\%; see Fig. \ref{fig:mtid}, left panel). 

In Fig. \ref{fig:mtid} we also show the evolution of the mass and half-mass radii for different numerical models. As we can see, for large enough particle numbers, the internal structures of the numerical models are quite similar; for example, the half-mass  radii of the systems are not dependent on the particle numbers of the object. 

We can conclude that the final models with $640k$ and $960k$ are practically identical in terms of both the mass loss and the internal mass distribution. As such, we find four to be the optimal number for the Ret-II dynamical system modelling (i.e. the $960k$ case). 

\begin{figure*}[htbp]
\centering
\includegraphics[width=0.45\linewidth]{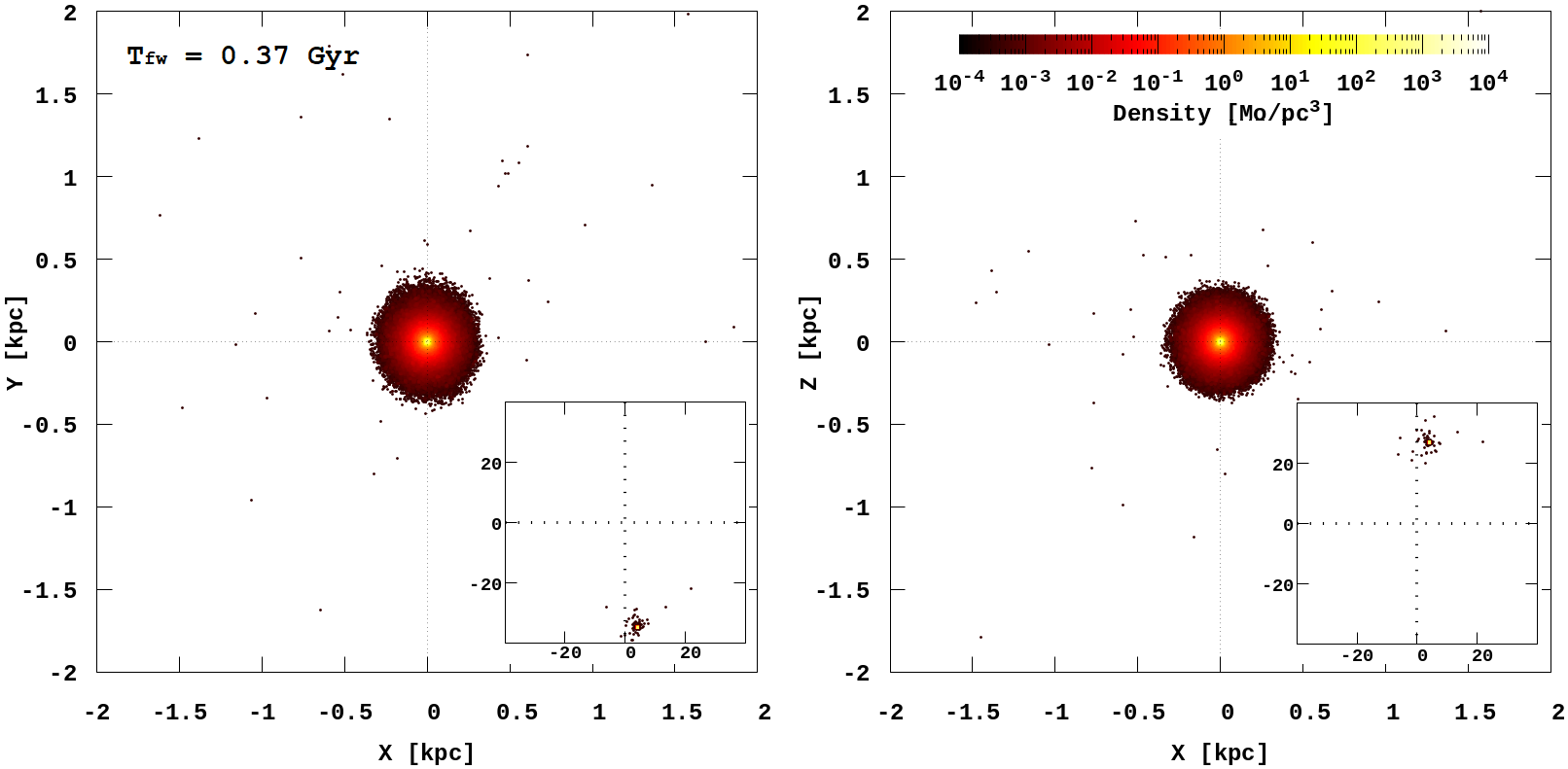}
\includegraphics[width=0.45\linewidth]{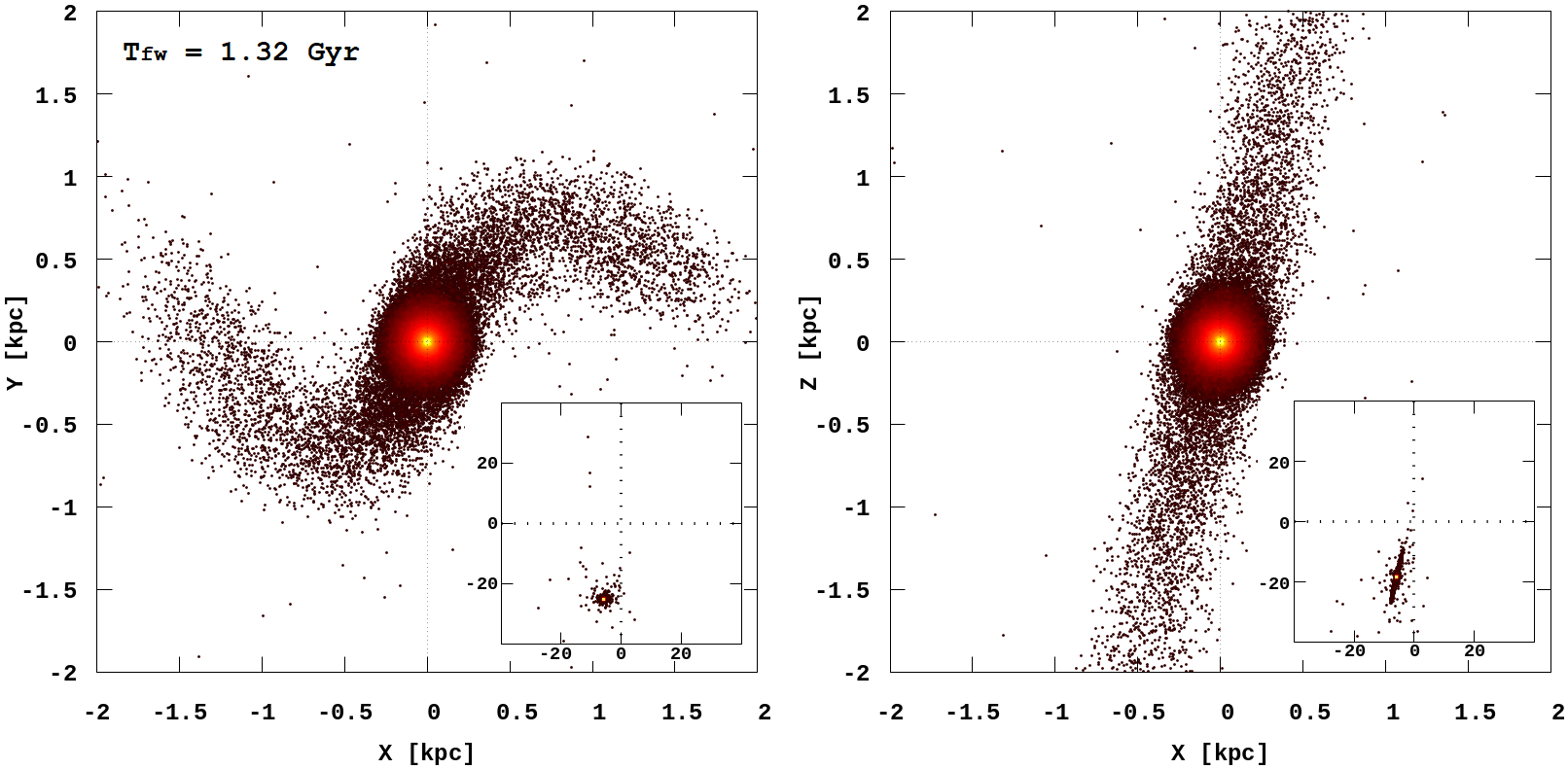}
\includegraphics[width=0.45\linewidth]{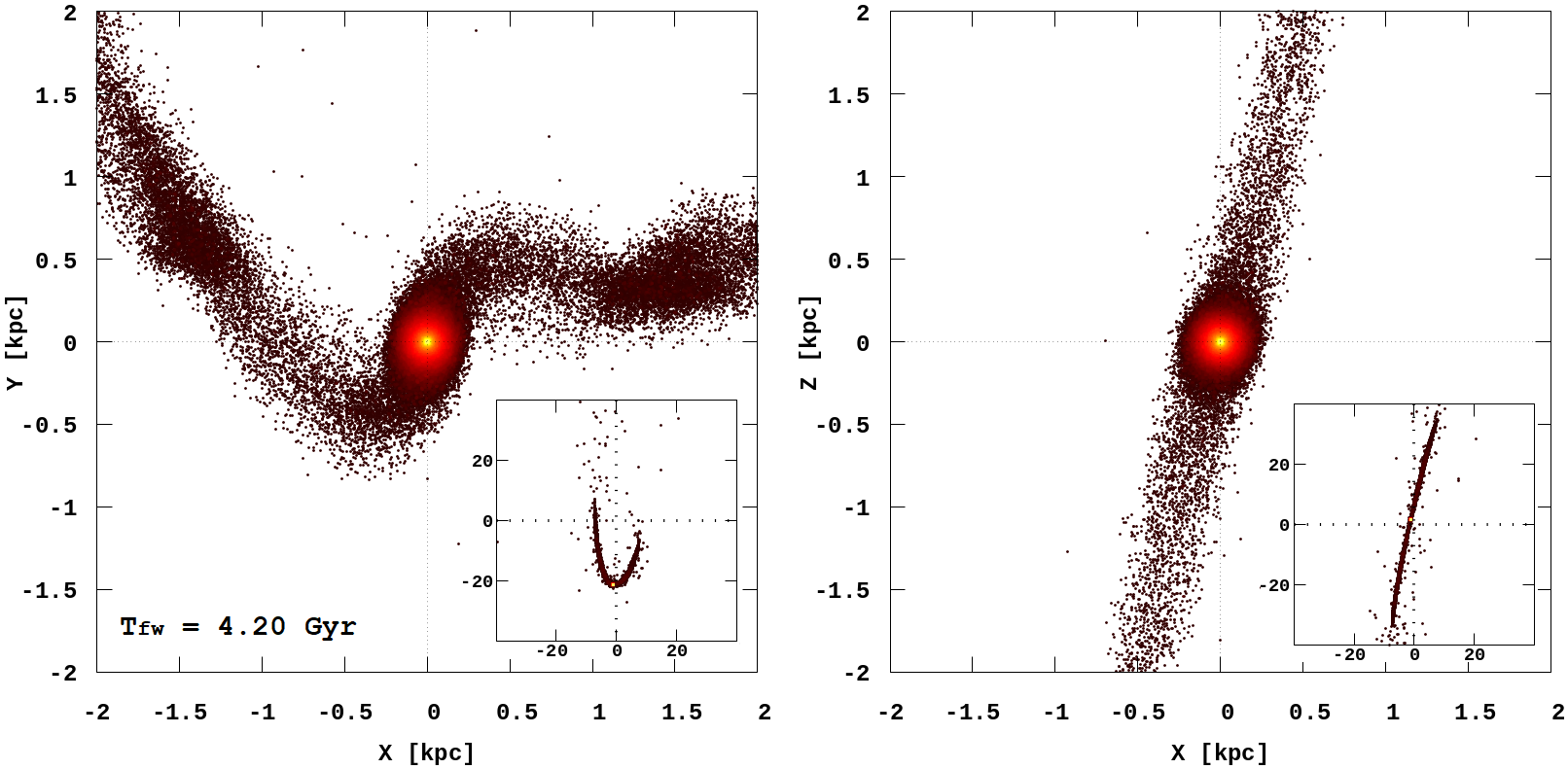}
\includegraphics[width=0.45\linewidth]{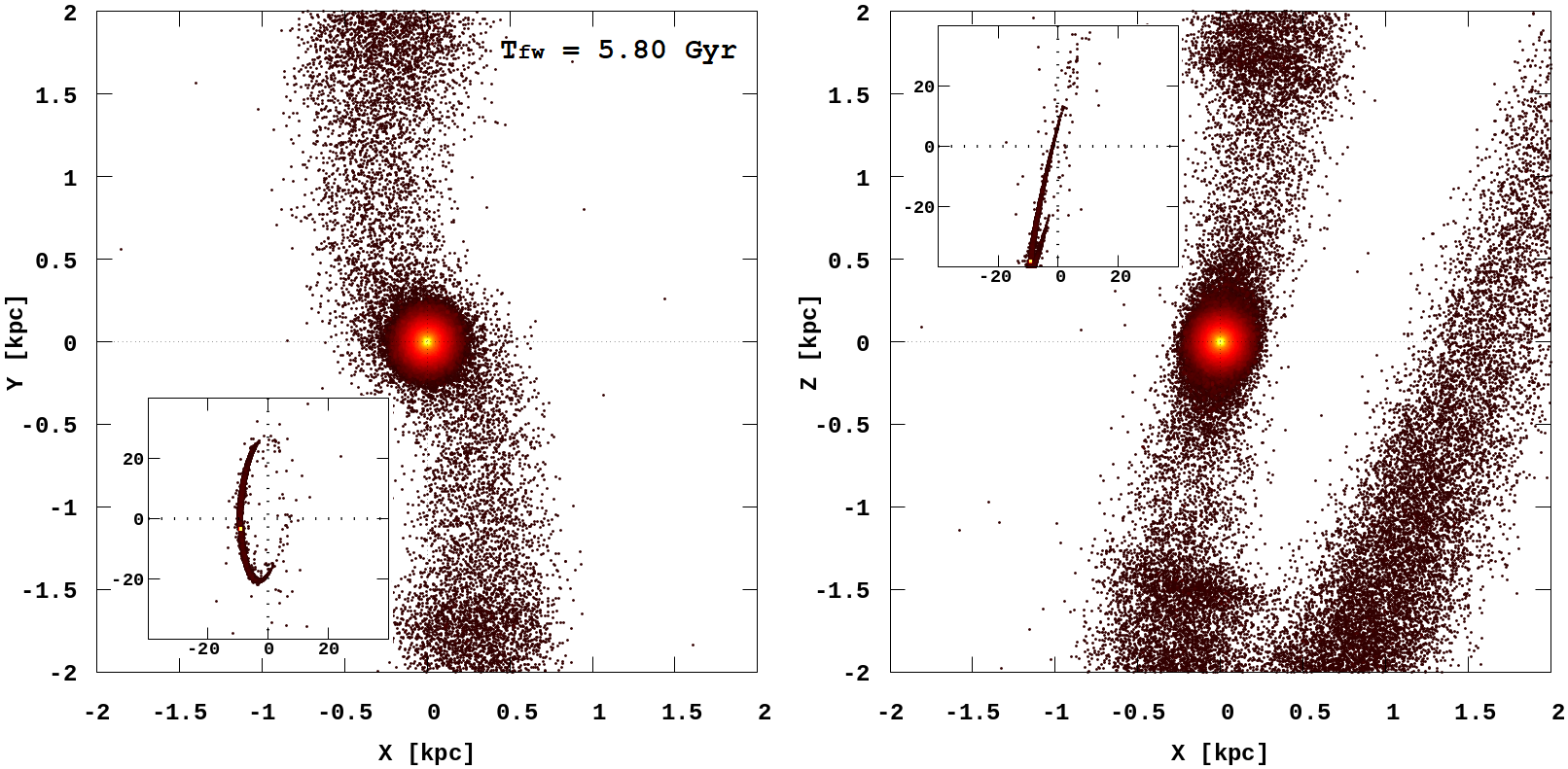}
\includegraphics[width=0.45\linewidth]{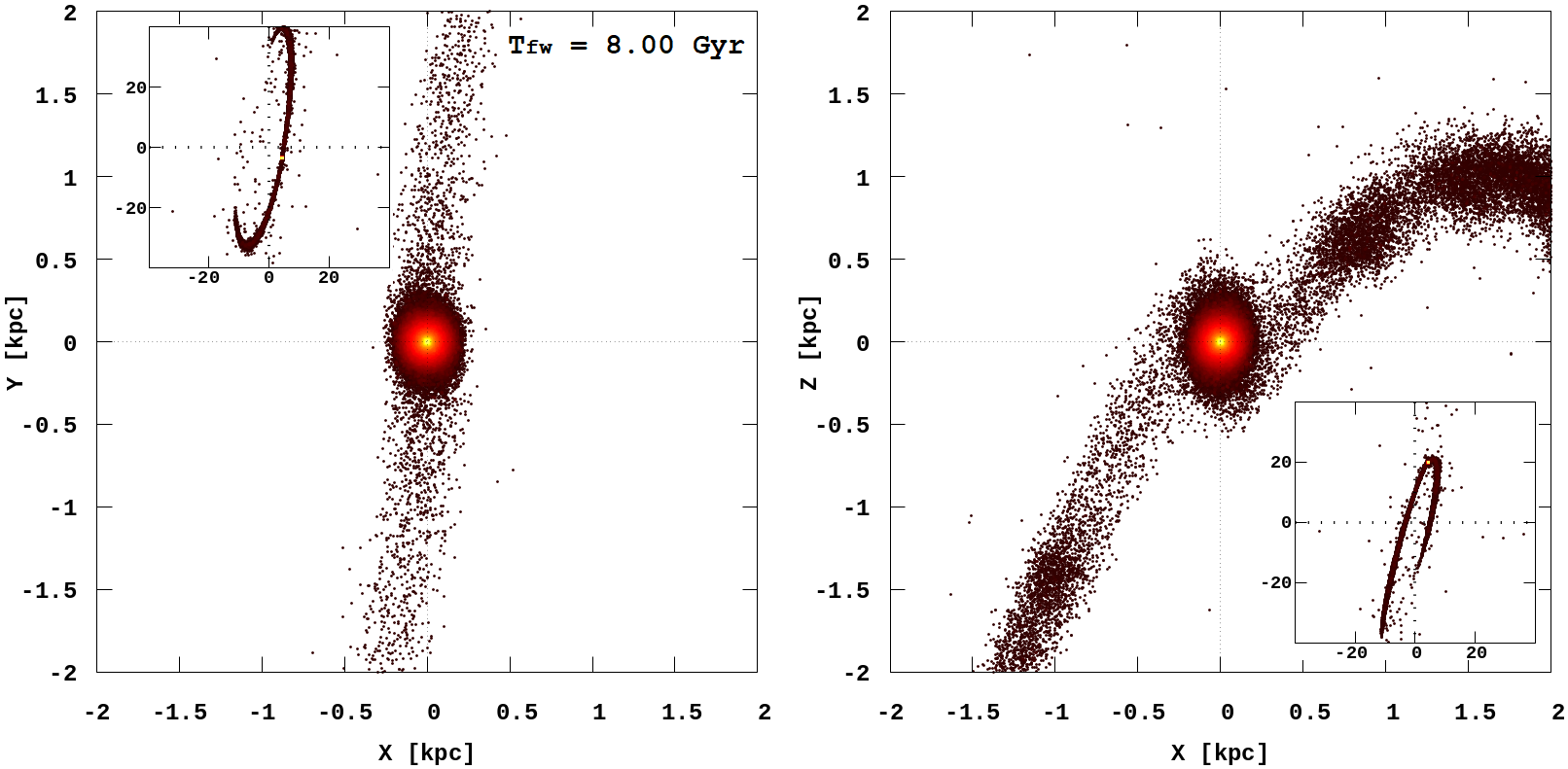}
\includegraphics[width=0.45\linewidth]{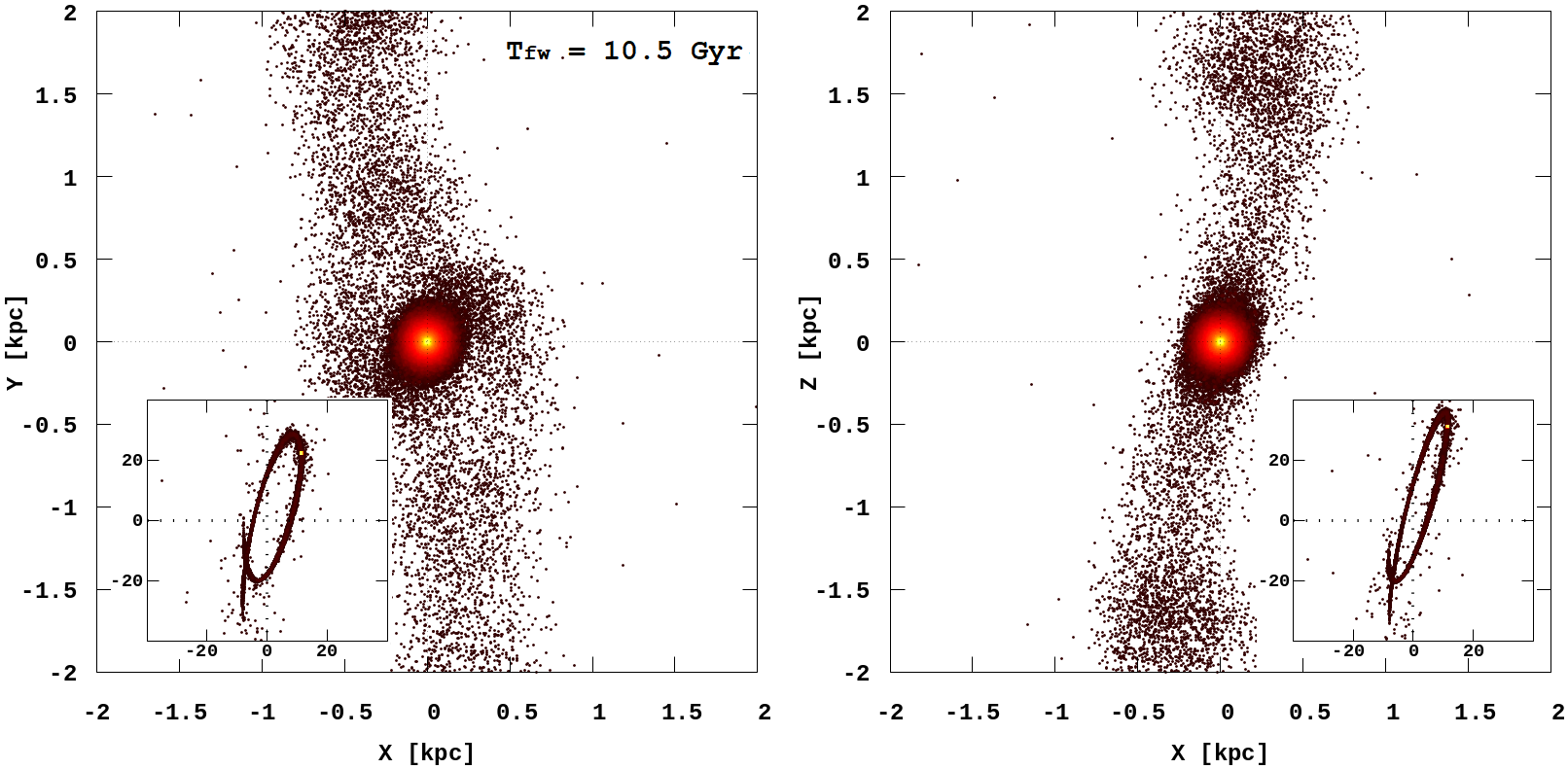}
\caption{Ret-II density distributions in {\tt 441327} TNG-TVP external potential for $N = 960k$. The central part in the local frame is presented in two projections, ($X$, $Y$) and ($X$, $Z$). The orbital global evolution is shown in the insets. The density distribution is presented for several moments of forward time integration: T = 0.37, 1.32, 4.20, 5.80, 8.00, and 10.5\, Gyr from \textit{left to right}.}
\label{fig:den}
\end{figure*}

In Fig. \ref{fig:den} we show the density distribution for Ret-II at different moments of time for the $N = 960k$ model. As we can see, the system has massive (many-kiloparsec-scale) extended tails. At a later stage, after 5\, Gyr, the escaped stars from the cluster fill almost the whole orbital volume of the UFD. Inside the tidal tails we see the prominent mass clumps generated by the complex Galactic orbital motion of Ret-II. The mass volume density inside these clumps easily reaches the 10$^{-2}$ M$_\odot$ / pc$^{3}$ level. Based on recent observations, the mass-to-light ratio of Ret-II inside its half-light radius is $\sim$470 M$_\odot$ / L$_\odot$, demonstrating that it is a strongly dark-matter-dominated UFD \citep{Simon2015, Walker2015, Mutlu-Pakdil2018}.

\section{Discussions and conclusions}\label{sec:phase-space}

\begin{figure*}[htbp]
\centering
\includegraphics[width=0.95\linewidth]{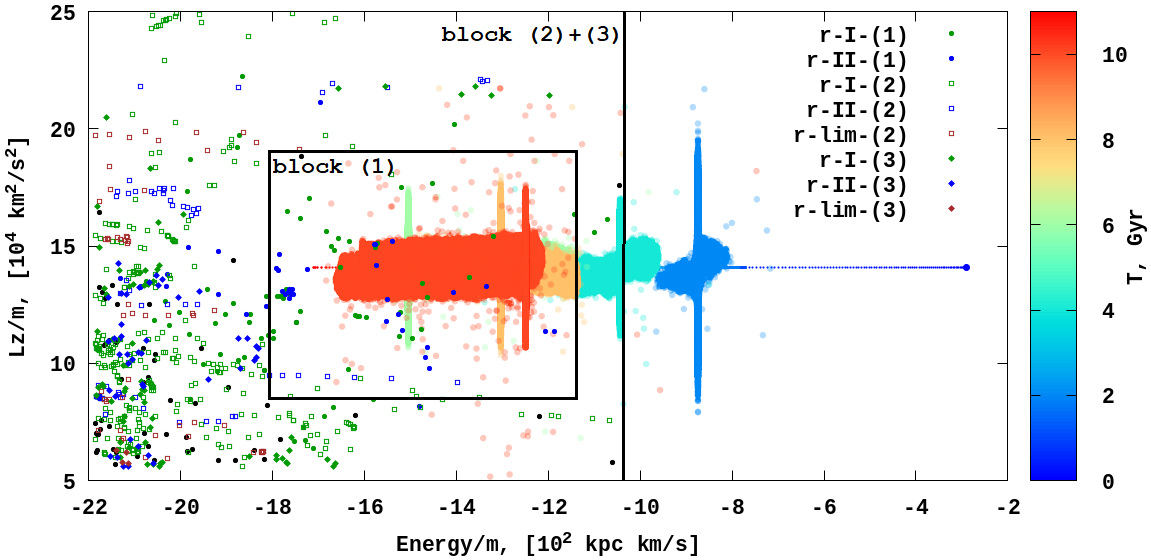}
\caption{Total specific energy ($E/m$) vs the $z$-th component of the specific angular momentum ($L_{\rm z}/m$) for the Ret-II and \textit{r} stars. The colours represent the time evolution of Ret-II from the past (blue) to today (red) based on a numerical simulation with N = 960k. The symbols (filled circle, unfilled square, and filled rhombus) represent the different types of \textit{r} stars (r-I, r-II, and r-lim). For each individual \textit{r} star, we generated ten random realisations inside their $\pm~\sigma$ error box. Block (1) comprises the stars presented in Table \ref{tab:star-group} as the `most probable' candidates, overlapping in the $E/m$ vs $L_{\rm z}/m$ phase-space with the present-day Ret-II. Block (2)+(3) represents the stars with the moderate `tentative' and low `in question' probabilities.}
\label{fig:ph-space}
\end{figure*}

Based on Fig. \ref{fig:den} we can conclude that in the energy angular momentum phase-space we definitely have a cross section of these escaped stars and the observed \textit{r}-PE stars. To show this more clearly, we plot the time evolution of the Ret-II members in specific energy ($E/m$) versus the \textit{z}-th component of the specific angular momentum ($L_{\rm z}/m$) space in Fig. \ref{fig:ph-space}. Here we present the specific energy and specific angular momentum of the stars for Ret-II. In the last\, gigayear of Ret-II's dynamical evolution, the phase-space distributions of particles for the tidal tails are almost in the same range. We also clearly see mixing in the phase-space of the observed \textit{r}-PE stars with our Ret-II escaped stars. 

As described in Sect. \ref{subsec:ini-r}, for each type of star we carried out ten initial data randomisations. Each of the observed \textit{r}-PE star positions and velocities was randomised and is shown in Fig. \ref{fig:ph-space}. During our $N$-body simulation, the objects from the Ret-II model spread across a large volume of phase-space. In the figure we only show a limited volume around the central values of the specific angular momentum $L_{\rm z}/m$ $\approx$ 150 000 km$^2$/s$^2$.   
 
We divided our results into three blocks (see Table \ref{tab:star-group} and Fig. \ref{fig:ph-space}). Blocks (2) and (3) are united in one common region, located to the left of the bold vertical line (at a specific energy of $\sim$-1000 kpc $\times$ km/s). Block (1) (-1800 kpc $\times$ km/s < $E/m$ < -1100 kpc $\times$ km/s and 80 000 km$^2$/s$^2$ < $L_{\rm z}/m$ < 190 000 km$^2$/s$^2$ ) contains the main region of the Ret-II stars from the numerical model.

Block (1) comprises the `most probable' stars, that is,  stars positioned inside or near the Ret-II object distribution (solid square in Fig.\ref{fig:ph-space}). In this area we find only two type of stars, \textbf{$r$-I} and \textbf{$r$-II}, and both types overlap significantly with the Ret-II particle cloud. To determine the confidence level of our results, we generated ten random realisations of each of the 530 \textit{r} stars (i.e. we have 5300 phase-space points). Depending on how many of these ten realisations hit the  phase-space area, we decided with which block we associate our individual stars. We present these probability values in Table A.1.  In total, we find 14 stars, and 10 of these 14 stars have a 100\% probability being former members of Ret-II. We estimated the probability based on our randomisation; a 100\% probability means that all ten random realisations of this individual \textit{r}-PE star hit the phase-space area of block (1). 

Block (2), marked as `tentative', comprises 54 stars and includes all three types of \textit{r} stars. All these  selected \textit{r}-PE stars have from a 70\% to a 100\% probability of being within the corresponding area of block (2)+(3). The phase-space area was selected based on the Ret-II stars from the numerical model, which are already beyond the main body of the dwarf galaxy.

In the last group, `in question', are 25 stars that have a probability of $\sim$40-60\% of being located in the same area of the phase-space region marked as block (2)+(3). This group of stars includes objects whose probability of belonging to the Ret-II progenitor is highly uncertain due to current measurement errors.

As a general conclusion, we can state that of the 530 known \textit{r} stars, at least 93 are former members of the Ret-II dynamical progenitor system. Our simple dynamical orbital modelling together with the chemical abundance analysis can in the future be  a powerful tool for such a peculiar star's progenitor identification in our Galaxy.  

Our investigation highlights the presence of a possible association between the\textit{r}-PE metal-poor stars in the Galactic halo and Ret-II. Notably, several Milky Way satellites, including Tucana\,III \citep{tucIII_rpe}, Grus\,II \citep{GrusII_rpe}, and Umi \citep{UMI_rpe}, also host stars exhibiting $r$-process enhancement. The recent discovery of an ultra-metal-poor star ([Fe/H] = $-4.15$) with [Eu/Fe] $<$ 1.04 in the Large Magellanic Cloud \citep{Chiti2024NatAs}  further complicates our understanding of the origin of\textit{ r}-PE stars. While we do not explore the potential associations of these \textit{r}-PE stars with the above-mentioned galaxies in our current work, our findings lay the groundwork for future research. In forthcoming publications, we plan to broaden our analysis by utilising a larger and updated sample that includes both Milky Way satellites and newly identified \textit{r}-PE stars.

\begin{acknowledgements}

The authors thank the anonymous referee for a very constructive report and suggestions that helped significantly improve the quality of the manuscript.

The authors gratefully acknowledge computing resources used for this modelling on the Gauss Centre for Supercomputing e.V. (GCS), through the John von Neumann Institute for Computing (NIC), with the GCS supercomputers JUWELS-booster at Julich Supercomputing Centre (JSC), Germany.

PB and OS thanks the support from the special program of the Polish Academy of Sciences and the U.S. National Academy of Sciences under the Long-term program to support Ukrainian research teams grant No.~PAN.BFB.S.BWZ.329.022.2023.

PB acknowledges the support by the National Science Foundation of China (NSFC) under grant No.~12473017.

M.K.M. acknowledges support from NSF grant OISE 1927130 (International Research Network for Nuclear Astrophysics/IReNA).

\end{acknowledgements}

\bibliographystyle{mnras}  
\bibliography{gc-mass-evol}   

\begin{thebibliography}{}
\makeatletter
\relax
\def\mn@urlcharsother{\let\do\@makeother \do\$\do\&\do\#\do\^\do\_\do\%\do\~}
\def\mn@doi{\begingroup\mn@urlcharsother \@ifnextchar [ {\mn@doi@} {\mn@doi@[]}}
\def\mn@doi@[#1]#2{\def\@tempa{#1}\ifx\@tempa\@empty \href {http://dx.doi.org/#2} {doi:#2}\else \href {http://dx.doi.org/#2} {#1}\fi \endgroup}
\def\mn@eprint#1#2{\mn@eprint@#1:#2::\@nil}
\def\mn@eprint@arXiv#1{\href {http://arxiv.org/abs/#1} {{\tt arXiv:#1}}}
\def\mn@eprint@dblp#1{\href {http://dblp.uni-trier.de/rec/bibtex/#1.xml} {dblp:#1}}
\def\mn@eprint@#1:#2:#3:#4\@nil{\def\@tempa {#1}\def\@tempb {#2}\def\@tempc {#3}\ifx \@tempc \@empty \let \@tempc \@tempb \let \@tempb \@tempa \fi \ifx \@tempb \@empty \def\@tempb {arXiv}\fi \@ifundefined {mn@eprint@\@tempb}{\@tempb:\@tempc}{\expandafter \expandafter \csname mn@eprint@\@tempb\endcsname \expandafter{\@tempc}}}

\bibitem[\protect\citeauthoryear{{Abohalima} \& {Frebel}}{{Abohalima} \& {Frebel}}{2018}]{jinabase}
{Abohalima} A.,  {Frebel} A.,  2018, \mn@doi [\apjs] {10.3847/1538-4365/aadfe9}, \href {https://ui.adsabs.harvard.edu/abs/2018ApJS..238...36A} {238, 36}

\bibitem[\protect\citeauthoryear{{Bechtol} et~al.,}{{Bechtol} et~al.}{2015}]{Bechtol2015}
{Bechtol} K.,  et~al., 2015, \mn@doi [\apj] {10.1088/0004-637X/807/1/50}, \href {https://ui.adsabs.harvard.edu/abs/2015ApJ...807...50B} {807, 50}

\bibitem[\protect\citeauthoryear{{Bennett} \& {Bovy}}{{Bennett} \& {Bovy}}{2019}]{Bennett2019}
{Bennett} M.,  {Bovy} J.,  2019, \mn@doi [\mnras] {10.1093/mnras/sty2813}, \href {https://ui.adsabs.harvard.edu/abs/2019MNRAS.482.1417B} {482, 1417}

\bibitem[\protect\citeauthoryear{{Bennett}, {Bovy}  \& {Hunt}}{{Bennett} et~al.}{2022}]{Bennett2022}
{Bennett} M.,  {Bovy} J.,   {Hunt} J. A.~S.,  2022, \mn@doi [\apj] {10.3847/1538-4357/ac5021}, \href {https://ui.adsabs.harvard.edu/abs/2022ApJ...927..131B} {927, 131}

\bibitem[\protect\citeauthoryear{{Benson}, {Frenk}, {Lacey}, {Baugh}  \& {Cole}}{{Benson} et~al.}{2002}]{Benson2002}
{Benson} A.~J.,  {Frenk} C.~S.,  {Lacey} C.~G.,  {Baugh} C.~M.,   {Cole} S.,  2002, \mn@doi [\mnras] {10.1046/j.1365-8711.2002.05388.x}, \href {https://ui.adsabs.harvard.edu/abs/2002MNRAS.333..177B} {333, 177}

\bibitem[\protect\citeauthoryear{{Berczik} et~al.,}{{Berczik} et~al.}{2011}]{Berczik2011}
{Berczik} P.,  et~al., 2011, in International conference on High Performance Computing, HPC-UA 2011. pp 8--18

\bibitem[\protect\citeauthoryear{{Berczik}, {Spurzem}  \& {Wang}}{{Berczik} et~al.}{2013}]{BSW2013}
{Berczik} P.,  {Spurzem} R.,   {Wang} L.,  2013, in Third International Conference on High Performance Computing, HPC-UA 2013. pp 52--59 (\mn@eprint {arXiv} {1312.1789})

\bibitem[\protect\citeauthoryear{{Bland-Hawthorn} \& {Gerhard}}{{Bland-Hawthorn} \& {Gerhard}}{2016}]{Bland-Hawthorn2016}
{Bland-Hawthorn} J.,  {Gerhard} O.,  2016, \mn@doi [\araa] {10.1146/annurev-astro-081915-023441}, \href {https://ui.adsabs.harvard.edu/abs/2016ARA&A..54..529B} {54, 529}

\bibitem[\protect\citeauthoryear{{Brauer}, {Ji}, {Drout}  \& {Frebel}}{{Brauer} et~al.}{2021}]{Brauer2021}
{Brauer} K.,  {Ji} A.~P.,  {Drout} M.~R.,   {Frebel} A.,  2021, \mn@doi [\apj] {10.3847/1538-4357/ac00b2}, \href {https://ui.adsabs.harvard.edu/abs/2021ApJ...915...81B} {915, 81}

\bibitem[\protect\citeauthoryear{{Brauer}, {Andales}, {Ji}, {Frebel}, {Mardini}, {G{\'o}mez}  \& {O'Shea}}{{Brauer} et~al.}{2022}]{Brauer2022}
{Brauer} K.,  {Andales} H.~D.,  {Ji} A.~P.,  {Frebel} A.,  {Mardini} M.~K.,  {G{\'o}mez} F.~A.,   {O'Shea} B.~W.,  2022, \mn@doi [\apj] {10.3847/1538-4357/ac85b9}, \href {https://ui.adsabs.harvard.edu/abs/2022ApJ...937...14B} {937, 14}

\bibitem[\protect\citeauthoryear{{Brown} et~al.,}{{Brown} et~al.}{2014}]{Brown2014}
{Brown} T.~M.,  et~al., 2014, \mn@doi [\apj] {10.1088/0004-637X/796/2/91}, \href {https://ui.adsabs.harvard.edu/abs/2014ApJ...796...91B} {796, 91}

\bibitem[\protect\citeauthoryear{{Chiti} et~al.,}{{Chiti} et~al.}{2023}]{Chiti2023AJ}
{Chiti} A.,  et~al., 2023, \mn@doi [\aj] {10.3847/1538-3881/aca416}, \href {https://ui.adsabs.harvard.edu/abs/2023AJ....165...55C} {165, 55}

\bibitem[\protect\citeauthoryear{{Chiti} et~al.,}{{Chiti} et~al.}{2024}]{Chiti2024NatAs}
{Chiti} A.,  et~al., 2024, \mn@doi [Nature Astronomy] {10.1038/s41550-024-02223-w}, \href {https://ui.adsabs.harvard.edu/abs/2024NatAs...8..637C} {8, 637}

\bibitem[\protect\citeauthoryear{{Cohen} \& {Huang}}{{Cohen} \& {Huang}}{2010}]{UMI_rpe}
{Cohen} J.~G.,  {Huang} W.,  2010, \mn@doi [\apj] {10.1088/0004-637X/719/1/931}, \href {https://ui.adsabs.harvard.edu/abs/2010ApJ...719..931C} {719, 931}

\bibitem[\protect\citeauthoryear{{Ernst}, {Just}, {Berczik}  \& {Olczak}}{{Ernst} et~al.}{2011}]{Ernst2011}
{Ernst} A.,  {Just} A.,  {Berczik} P.,   {Olczak} C.,  2011, \mn@doi [\aap] {10.1051/0004-6361/201118021}, \href {https://ui.adsabs.harvard.edu/abs/2011A&A...536A..64E} {536, A64}

\bibitem[\protect\citeauthoryear{{Frebel}}{{Frebel}}{2018}]{Frebel2018ARNPS}
{Frebel} A.,  2018, \mn@doi [Annual Review of Nuclear and Particle Science] {10.1146/annurev-nucl-101917-021141}, \href {https://ui.adsabs.harvard.edu/abs/2018ARNPS..68..237F} {68, 237}

\bibitem[\protect\citeauthoryear{{Frebel}}{{Frebel}}{2019}]{Frebel2019}
{Frebel} A.,  2019, \mn@doi [Annals of Physics] {10.1016/j.aop.2019.167909}, \href {https://ui.adsabs.harvard.edu/abs/2019AnPhy.41067909F} {410, 167909}

\bibitem[\protect\citeauthoryear{{Frebel} \& {Norris}}{{Frebel} \& {Norris}}{2015}]{Frebel2015}
{Frebel} A.,  {Norris} J.~E.,  2015, \mn@doi [\araa] {10.1146/annurev-astro-082214-122423}, \href {https://ui.adsabs.harvard.edu/abs/2015ARA&A..53..631F} {53, 631}

\bibitem[\protect\citeauthoryear{{Frebel}, {Kirby}  \& {Simon}}{{Frebel} et~al.}{2010a}]{Frebel_2010_nat}
{Frebel} A.,  {Kirby} E.~N.,   {Simon} J.~D.,  2010a, \mn@doi [\nat] {10.1038/nature08772}, \href {https://ui.adsabs.harvard.edu/abs/2010Natur.464...72F} {464, 72}

\bibitem[\protect\citeauthoryear{{Frebel}, {Simon}, {Geha}  \& {Willman}}{{Frebel} et~al.}{2010b}]{Frebel2010}
{Frebel} A.,  {Simon} J.~D.,  {Geha} M.,   {Willman} B.,  2010b, \mn@doi [\apj] {10.1088/0004-637X/708/1/560}, \href {https://ui.adsabs.harvard.edu/abs/2010ApJ...708..560F} {708, 560}

\bibitem[\protect\citeauthoryear{{Gaia Collaboration} et~al.,}{{Gaia Collaboration} et~al.}{2016}]{Gaia_the_mission}
{Gaia Collaboration} et~al., 2016, \mn@doi [\aap] {10.1051/0004-6361/201629272}, \href {https://ui.adsabs.harvard.edu/abs/2016A&A...595A...1G} {595, A1}

\bibitem[\protect\citeauthoryear{{Gaia Collaboration} et~al.,}{{Gaia Collaboration} et~al.}{2023}]{Gaia_DR3}
{Gaia Collaboration} et~al., 2023, \mn@doi [\aap] {10.1051/0004-6361/202243940}, \href {https://ui.adsabs.harvard.edu/abs/2023A&A...674A...1G} {674, A1}

\bibitem[\protect\citeauthoryear{{Geha}, {Willman}, {Simon}, {Strigari}, {Kirby}, {Law}  \& {Strader}}{{Geha} et~al.}{2009}]{Geha2009}
{Geha} M.,  {Willman} B.,  {Simon} J.~D.,  {Strigari} L.~E.,  {Kirby} E.~N.,  {Law} D.~R.,   {Strader} J.,  2009, \mn@doi [\apj] {10.1088/0004-637X/692/2/1464}, \href {https://ui.adsabs.harvard.edu/abs/2009ApJ...692.1464G} {692, 1464}

\bibitem[\protect\citeauthoryear{{Gravity Collaboration} et~al.,}{{Gravity Collaboration} et~al.}{2019}]{Gravity2019}
{Gravity Collaboration} et~al., 2019, \mn@doi [\aap] {10.1051/0004-6361/201935656}, \href {https://ui.adsabs.harvard.edu/abs/2019A&A...625L..10G} {625, L10}

\bibitem[\protect\citeauthoryear{{Hansen} et~al.,}{{Hansen} et~al.}{2017}]{tucIII_rpe}
{Hansen} T.~T.,  et~al., 2017, \mn@doi [\apj] {10.3847/1538-4357/aa634a}, \href {https://ui.adsabs.harvard.edu/abs/2017ApJ...838...44H} {838, 44}

\bibitem[\protect\citeauthoryear{{Hansen} et~al.,}{{Hansen} et~al.}{2020}]{GrusII_rpe}
{Hansen} T.~T.,  et~al., 2020, \mn@doi [\apj] {10.3847/1538-4357/ab9643}, \href {https://ui.adsabs.harvard.edu/abs/2020ApJ...897..183H} {897, 183}

\bibitem[\protect\citeauthoryear{{Ishchenko}, {Sobolenko}, {Berczik}, {Khoperskov}, {Omarov}, {Sobodar}  \& {Makukov}}{{Ishchenko} et~al.}{2023}]{Ishchenko2023a}
{Ishchenko} M.,  {Sobolenko} M.,  {Berczik} P.,  {Khoperskov} S.,  {Omarov} C.,  {Sobodar} O.,   {Makukov} M.,  2023, \mn@doi [\aap] {10.1051/0004-6361/202245117}, \href {https://ui.adsabs.harvard.edu/abs/2023A&A...673A.152I} {673, A152}

\bibitem[\protect\citeauthoryear{{Ishchenko} et~al.,}{{Ishchenko} et~al.}{2024}]{Ishchenko2024mass-loss}
{Ishchenko} M.,  et~al., 2024, \mn@doi [\aap] {10.1051/0004-6361/202450399}, \href {https://ui.adsabs.harvard.edu/abs/2024A&A...689A.178I} {689, A178}

\bibitem[\protect\citeauthoryear{{Ji}, {Simon}, {Frebel}, {Venn}  \& {Hansen}}{{Ji} et~al.}{2019}]{Ji2019}
{Ji} A.~P.,  {Simon} J.~D.,  {Frebel} A.,  {Venn} K.~A.,   {Hansen} T.~T.,  2019, \mn@doi [\apj] {10.3847/1538-4357/aaf3bb}, \href {https://ui.adsabs.harvard.edu/abs/2019ApJ...870...83J} {870, 83}

\bibitem[\protect\citeauthoryear{{Ji} et~al.,}{{Ji} et~al.}{2023}]{Ji2023}
{Ji} A.~P.,  et~al., 2023, \mn@doi [\aj] {10.3847/1538-3881/acad84}, \href {https://ui.adsabs.harvard.edu/abs/2023AJ....165..100J} {165, 100}

\bibitem[\protect\citeauthoryear{{Johnson} \& {Soderblom}}{{Johnson} \& {Soderblom}}{1987}]{Johnson1987}
{Johnson} D. R.~H.,  {Soderblom} D.~R.,  1987, \mn@doi [\aj] {10.1086/114370}, \href {https://ui.adsabs.harvard.edu/abs/1987AJ.....93..864J} {93, 864}

\bibitem[\protect\citeauthoryear{{Just}, {Berczik}, {Petrov}  \& {Ernst}}{{Just} et~al.}{2009}]{Just2009}
{Just} A.,  {Berczik} P.,  {Petrov} M.~I.,   {Ernst} A.,  2009, \mn@doi [\mnras] {10.1111/j.1365-2966.2008.14099.x}, \href {https://ui.adsabs.harvard.edu/abs/2009MNRAS.392..969J} {392, 969}

\bibitem[\protect\citeauthoryear{{Karim} \& {Mamajek}}{{Karim} \& {Mamajek}}{2017}]{Karim2017}
{Karim} T.,  {Mamajek} E.~E.,  2017, \mn@doi [\mnras] {10.1093/mnras/stw2772}, \href {https://ui.adsabs.harvard.edu/abs/2017MNRAS.465..472K} {465, 472}

\bibitem[\protect\citeauthoryear{{King}}{{King}}{1962}]{King1962}
{King} I.,  1962, \mn@doi [\aj] {10.1086/108756}, \href {https://ui.adsabs.harvard.edu/abs/1962AJ.....67..471K} {67, 471}

\bibitem[\protect\citeauthoryear{{Koch}, {Feltzing}, {Ad{\'e}n}  \& {Matteucci}}{{Koch} et~al.}{2013}]{Koch2013}
{Koch} A.,  {Feltzing} S.,  {Ad{\'e}n} D.,   {Matteucci} F.,  2013, \mn@doi [\aap] {10.1051/0004-6361/201220742}, \href {https://ui.adsabs.harvard.edu/abs/2013A&A...554A...5K} {554, A5}

\bibitem[\protect\citeauthoryear{{Koposov}, {Belokurov}, {Torrealba}  \& {Evans}}{{Koposov} et~al.}{2015}]{Koposov2015}
{Koposov} S.~E.,  {Belokurov} V.,  {Torrealba} G.,   {Evans} N.~W.,  2015, \mn@doi [\apj] {10.1088/0004-637X/805/2/130}, \href {https://ui.adsabs.harvard.edu/abs/2015ApJ...805..130K} {805, 130}

\bibitem[\protect\citeauthoryear{{Lindegren} et~al.,}{{Lindegren} et~al.}{2021}]{Lindegren_Parallax_2021}
{Lindegren} L.,  et~al., 2021, \mn@doi [\aap] {10.1051/0004-6361/202039653}, \href {https://ui.adsabs.harvard.edu/abs/2021A&A...649A...4L} {649, A4}

\bibitem[\protect\citeauthoryear{{Makino} \& {Aarseth}}{{Makino} \& {Aarseth}}{1992}]{MA1992}
{Makino} J.,  {Aarseth} S.~J.,  1992, \pasj, \href {https://ui.adsabs.harvard.edu/abs/1992PASJ...44..141M} {44, 141}

\bibitem[\protect\citeauthoryear{{Mardini} et~al.,}{{Mardini} et~al.}{2020}]{Mardini2020}
{Mardini} M.~K.,  et~al., 2020, \mn@doi [\apj] {10.3847/1538-4357/abbc13}, \href {https://ui.adsabs.harvard.edu/abs/2020ApJ...903...88M} {903, 88}

\bibitem[\protect\citeauthoryear{{Mardini} et~al.,}{{Mardini} et~al.}{2022a}]{Mardini2022_j1808}
{Mardini} M.~K.,  et~al., 2022a, \mn@doi [\mnras] {10.1093/mnras/stac2783}, \href {https://ui.adsabs.harvard.edu/abs/2022MNRAS.517.3993M} {517, 3993}

\bibitem[\protect\citeauthoryear{{Mardini}, {Frebel}, {Chiti}, {Meiron}, {Brauer}  \& {Ou}}{{Mardini} et~al.}{2022b}]{Mardini2022Atari}
{Mardini} M.~K.,  {Frebel} A.,  {Chiti} A.,  {Meiron} Y.,  {Brauer} K.~V.,   {Ou} X.,  2022b, \mn@doi [\apj] {10.3847/1538-4357/ac8102}, \href {https://ui.adsabs.harvard.edu/abs/2022ApJ...936...78M} {936, 78}

\bibitem[\protect\citeauthoryear{{Mardini}, {Frebel}, {Betre}, {Jacobson}, {Norris}  \& {Christlieb}}{{Mardini} et~al.}{2024a}]{Mardini2024}
{Mardini} M.~K.,  {Frebel} A.,  {Betre} L.,  {Jacobson} H.,  {Norris} J.~E.,   {Christlieb} N.,  2024a, \mn@doi [\mnras] {10.1093/mnras/stad3925}, \href {https://ui.adsabs.harvard.edu/abs/2024MNRAS.528.2912M} {528, 2912}

\bibitem[\protect\citeauthoryear{{Mardini}, {Frebel}  \& {Chiti}}{{Mardini} et~al.}{2024b}]{Mardini2024ump}
{Mardini} M.~K.,  {Frebel} A.,   {Chiti} A.,  2024b, \mn@doi [\mnras] {10.1093/mnrasl/slad197}, \href {https://ui.adsabs.harvard.edu/abs/2024MNRAS.529L..60M} {529, L60}

\bibitem[\protect\citeauthoryear{{Martin}, {Ibata}, {Chapman}, {Irwin}  \& {Lewis}}{{Martin} et~al.}{2007}]{Martin_2007}
{Martin} N.~F.,  {Ibata} R.~A.,  {Chapman} S.~C.,  {Irwin} M.,   {Lewis} G.~F.,  2007, \mn@doi [\mnras] {10.1111/j.1365-2966.2007.12055.x}, \href {https://ui.adsabs.harvard.edu/abs/2007MNRAS.380..281M} {380, 281}

\bibitem[\protect\citeauthoryear{{Minor}, {Pace}, {Marshall}  \& {Strigari}}{{Minor} et~al.}{2019}]{Minor2019}
{Minor} Q.~E.,  {Pace} A.~B.,  {Marshall} J.~L.,   {Strigari} L.~E.,  2019, \mn@doi [\mnras] {10.1093/mnras/stz1468}, \href {https://ui.adsabs.harvard.edu/abs/2019MNRAS.487.2961M} {487, 2961}

\bibitem[\protect\citeauthoryear{{Mutlu-Pakdil} et~al.,}{{Mutlu-Pakdil} et~al.}{2018}]{Mutlu-Pakdil2018}
{Mutlu-Pakdil} B.,  et~al., 2018, \mn@doi [\apj] {10.3847/1538-4357/aacd0e}, \href {https://ui.adsabs.harvard.edu/abs/2018ApJ...863...25M} {863, 25}

\bibitem[\protect\citeauthoryear{{Nelson} et~al.,}{{Nelson} et~al.}{2018}]{Nelson2018}
{Nelson} D.,  et~al., 2018, \mn@doi [\mnras] {10.1093/mnras/stx3040}, \href {https://ui.adsabs.harvard.edu/abs/2018MNRAS.475..624N} {475, 624}

\bibitem[\protect\citeauthoryear{{Nelson} et~al.,}{{Nelson} et~al.}{2019a}]{Nelson2019}
{Nelson} D.,  et~al., 2019a, \mn@doi [Computational Astrophysics and Cosmology] {10.1186/s40668-019-0028-x}, \href {https://ui.adsabs.harvard.edu/abs/2019ComAC...6....2N} {6, 2}

\bibitem[\protect\citeauthoryear{{Nelson} et~al.,}{{Nelson} et~al.}{2019b}]{NelsonPill2019}
{Nelson} D.,  et~al., 2019b, \mn@doi [\mnras] {10.1093/mnras/stz2306}, \href {https://ui.adsabs.harvard.edu/abs/2019MNRAS.490.3234N} {490, 3234}

\bibitem[\protect\citeauthoryear{{Nitadori} \& {Makino}}{{Nitadori} \& {Makino}}{2008}]{Nitadori2008}
{Nitadori} K.,  {Makino} J.,  2008, \mn@doi [\na] {10.1016/j.newast.2008.01.010}, \href {https://ui.adsabs.harvard.edu/abs/2008NewA...13..498N} {13, 498}

\bibitem[\protect\citeauthoryear{{Ou} et~al.,}{{Ou} et~al.}{2024}]{Xiaowei2024}
{Ou} X.,  et~al., 2024, \mn@doi [\apj] {10.3847/1538-4357/ad2f27}, \href {https://ui.adsabs.harvard.edu/abs/2024ApJ...966...33O} {966, 33}

\bibitem[\protect\citeauthoryear{{Reid} \& {Brunthaler}}{{Reid} \& {Brunthaler}}{2004}]{Reid2004}
{Reid} M.~J.,  {Brunthaler} A.,  2004, \mn@doi [\apj] {10.1086/424960}, \href {https://ui.adsabs.harvard.edu/abs/2004ApJ...616..872R} {616, 872}

\bibitem[\protect\citeauthoryear{{Roederer} et~al.,}{{Roederer} et~al.}{2016}]{Roederer2016}
{Roederer} I.~U.,  et~al., 2016, \mn@doi [\aj] {10.3847/0004-6256/151/3/82}, \href {https://ui.adsabs.harvard.edu/abs/2016AJ....151...82R} {151, 82}

\bibitem[\protect\citeauthoryear{{Sch{\"o}nrich}, {Binney}  \& {Dehnen}}{{Sch{\"o}nrich} et~al.}{2010}]{Schonrich2010}
{Sch{\"o}nrich} R.,  {Binney} J.,   {Dehnen} W.,  2010, \mn@doi [\mnras] {10.1111/j.1365-2966.2010.16253.x}, \href {https://ui.adsabs.harvard.edu/abs/2010MNRAS.403.1829S} {403, 1829}

\bibitem[\protect\citeauthoryear{{Simon} et~al.,}{{Simon} et~al.}{2015}]{Simon2015}
{Simon} J.~D.,  et~al., 2015, \mn@doi [\apj] {10.1088/0004-637X/808/1/95}, \href {https://ui.adsabs.harvard.edu/abs/2015ApJ...808...95S} {808, 95}

\bibitem[\protect\citeauthoryear{{Simon} et~al.,}{{Simon} et~al.}{2023}]{Simon2023}
{Simon} J.~D.,  et~al., 2023, \mn@doi [\apj] {10.3847/1538-4357/aca9d1}, 944, 43

\bibitem[\protect\citeauthoryear{{Walker}, {Mateo}, {Olszewski}, {Pe{\~n}arrubia}, {Evans}  \& {Gilmore}}{{Walker} et~al.}{2009}]{Walker2009}
{Walker} M.~G.,  {Mateo} M.,  {Olszewski} E.~W.,  {Pe{\~n}arrubia} J.,  {Evans} N.~W.,   {Gilmore} G.,  2009, \mn@doi [\apj] {10.1088/0004-637X/704/2/1274}, \href {https://ui.adsabs.harvard.edu/abs/2009ApJ...704.1274W} {704, 1274}

\bibitem[\protect\citeauthoryear{{Walker}, {Mateo}, {Olszewski}, {Bailey}, {Koposov}, {Belokurov}  \& {Evans}}{{Walker} et~al.}{2015}]{Walker2015}
{Walker} M.~G.,  {Mateo} M.,  {Olszewski} E.~W.,  {Bailey} John~I. I.,  {Koposov} S.~E.,  {Belokurov} V.,   {Evans} N.~W.,  2015, \mn@doi [\apj] {10.1088/0004-637X/808/2/108}, \href {https://ui.adsabs.harvard.edu/abs/2015ApJ...808..108W} {808, 108}

\bibitem[\protect\citeauthoryear{{Wolf}, {Martinez}, {Bullock}, {Kaplinghat}, {Geha}, {Mu{\~n}oz}, {Simon}  \& {Avedo}}{{Wolf} et~al.}{2010}]{Wolf2010}
{Wolf} J.,  {Martinez} G.~D.,  {Bullock} J.~S.,  {Kaplinghat} M.,  {Geha} M.,  {Mu{\~n}oz} R.~R.,  {Simon} J.~D.,   {Avedo} F.~F.,  2010, \mn@doi [\mnras] {10.1111/j.1365-2966.2010.16753.x}, \href {https://ui.adsabs.harvard.edu/abs/2010MNRAS.406.1220W} {406, 1220}

\bibitem[\protect\citeauthoryear{{van Donkelaar}, {Mayer}, {Capelo}, {Tamfal}, {Quinn}  \& {Madau}}{{van Donkelaar} et~al.}{2023}]{Mayer2023}
{van Donkelaar} F.,  {Mayer} L.,  {Capelo} P.~R.,  {Tamfal} T.,  {Quinn} T.~R.,   {Madau} P.,  2023, \mn@doi [\mnras] {10.1093/mnras/stad946}, \href {https://ui.adsabs.harvard.edu/abs/2023MNRAS.522.1726V} {522, 1726}

\bibitem[\protect\citeauthoryear{{van Donkelaar}, {Mayer}, {Capelo}, {Tamfal}, {Quinn}  \& {Madau}}{{van Donkelaar} et~al.}{2024}]{Mayer2024}
{van Donkelaar} F.,  {Mayer} L.,  {Capelo} P.~R.,  {Tamfal} T.,  {Quinn} T.~R.,   {Madau} P.,  2024, \mn@doi [\mnras] {10.1093/mnras/stae804}, \href {https://ui.adsabs.harvard.edu/abs/2024MNRAS.529.4104V} {529, 4104}

\makeatother
\end{thebibliography}

\begin{appendix}

\onecolumn

\section{Various \textit{r}-PE groups detected in our study along with Ret-II membership likelihood.}\label{app:ret}

\begin{table*}[h!]
\setlength{\tabcolsep}{2pt}
\centering
\caption{Various \textit{r}-PE groups detected in our study along with their Ret-II membership likelihood.}
\label{tab:star-group}
\begin{tabular}{lcclcclcc}
\hline
\hline 
ID & $\%$ & Type & ID & $\%$ & Type &  ID & $\%$ & Type \\
\hline
\hline
\textbf{Most probable -- (block 1)} &   & &  &  &  &  &  &  \\
RAVEJ183013.5-455510 & 100 & r-I & 2MASSJ14232679-2834200 & 100 & r-I & MASSJ05241392-0336543 & 100 & r-II \\ 
HE0420+0123a  & 100  & r-II & 2MASSJ04315411-0632100 & 60 & r-I & 2MASSJ03154102-7626329 & 100 & r-I \\
2MASSJ07501424-4123454 &  50  & r-I & HD106373 & 100 & r-I & SMSSJ183647.89-274333.1 & 100 & r-II \\ 
2MASSJ00482431-1041309 &  100 & r-I & 2MASSJ08594093-1415151 & 100 & r-I & 2MASSJ02031860-7930291 & 70 & r-I \\ 
2MASSJ13374885-0826176 & 100  & r-II & 2MASSJ12341308-3149577 & 80 & r-I & & & \\ 
\hline
\textbf{Tentative -- (block 2)}   &   &  &  &  & &  &  &  \\ 
2MASSJ00452379-2112161 & 90  & r-I & E0045-2430 & 90 & r-I & 2MASSJ02070641-5009166 & 80 & r-I \\ 
HD13979      &  90 & r-lim & 2MASSJ02462013-1518419 & 80 & r-II & BPSCS31078-018 & 100 & r-II \\ 
2MASSJ03210882-3944213 & 100  & r-I & CD-241782 & 80 & r-lim & 2MASSJ03550926-0637108 & 80 & r-lim \\ 
MASSJ03563703-5838281  & 70  & r-lim & 2MASSJ04192966-0517491 & 70 & r-I & BPSCS22186-025 & 70 & r-I \\ 
HIP22246 &  100 & r-I & HE0516-3820 & 10 & r-I & HE0524-2055 & 70 & r-I \\
2MASSJ05311779-5810048 &  70 & r-I & 2MASSJ05384334-5147228 & 100 & r-lim & 2MASSJ06195001-5312114 & 100 & r-II \\
2MASSJ06332771-3519240 &  100 & r-I & 2MASSJ07052028-3343242 & 100 & r-I & 2MASSJ08393460-2122069 & 90 & r-I\\
2MASSJ10063414-7030212  & 80  & r-I & 2MASSJ10362687-3746174 & 100 & r-II & HE1131+0141 & 70 & r-II \\
2MASSJ11404944-1615396 &  70 & r-II & RAVEJ115941.7-382043  & 70 & r-II & HD107752 & 80 & r-I \\
2MASSJ12255123-2351074 &  100 & r-I & 2MASSJ12292696-0442325 & 100 & r-I & 2MASSJ13261792-0945176 & 90 & r-I \\
2MASSJ14101587-0343553 &  70 & r-I & HE1429-0347 & 100 & r-I & 2MASSJ14435196-2106283 & 80 & r-lim \\
2MASSJ14592981-3852558 &  100 & r-II & 2MASSJ15002498-0613374 & 80 & r-I & 2MASSJ15062866-1428038 & 80 & r-I \\
2MASSJ15211026-0607566 &  90 & r-II & 2MASSJ15582962-1224344 & 100 & r-I & SMSSJ160447.75-293146.7 & 80 & r-I \\
HD149414 &  100 & r-I & HD175305 & 90  & r-I & 2MASSJ18562774-7251331 & 100 & r-I \\
2MASSJ20504869-3355289 &  70 & r-I & 2MASSJ21370807-0927347 & 70 & r-lim & 2MASSJ21462219-0512324 & 100 & r-I \\
G214-1 & 80  & r-I & BPSCS22875-029 & 70 & r-II & HE2229-4153 & 70 & r-I \\
HE2242-1930 &  70 & r-I & 2MASSJ23060975-6107368 & 70 & r-I & 2MASSJ23242766-2728055  & 80 & r-I \\
2MASSJ23265258-0159248 &  80 & r-I &  MASSJ23411581-6406440 & 10 & r-I & 2MASSJ23490902-2447176 & 90 & r-I \\
\hline
\textbf{In question -- (block 3)} &   & &  &  &  &  &  &  \\
HD224930 & 40  & r-I & BPSCS22957-036 & 60 & r-I & BPSCS22958-037 & 66 & r-I \\
HE0240-0807 &  40 & r-II & 2MASSJ02570028-3024374 & 50 & r-I & 2MASSJ03133726-1020553 & 50 & r-I \\
2MASSJ03422816-6500355 & 60  & r-II & 2MASSJ05381700-7516207 & 50 & r-I & HE0538-4515 & 50 & r-II \\
SMSSJ062609.83-590503.2 &  60 & r-II & 2MASSJ08015897-5752032 & 50 & r-I & 2MASSJ09261133-1526232 & 50 & r-I \\
2MASSJ09544277+5246414 & 50  & r-II & 2MASSJ10344785-4823544 & 40 & r-lim & SMSSJ105320.99-435300.1 & 50 & r-I \\
HE1127-1143 & 60  & r-II & 2MASSJ11301705-1449325 & 40 & r-I & BPSBS16083-172 & 50 & r-II \\
PSCS30306-132 &  40 & r-II & BPSCS22878-121 & 40 & r-I & 2MASSJ19552158-4613569 & 60 & r-I \\
2MASSJ22163596+0246171 & 50  & r-I & BPSCS22888-047 & 40 & r-II & MASSJ23362202-5607498 & 60 & r-II \\
2MASSJ19232518-5833410 & 40 & r-II  &  &  &  & &  &  \\
\hline 
\end{tabular}
\tablefoot{This table is based on the associations that are visible in Fig. \ref{fig:ph-space}. The detailed properties of these groups are described in Sect. \ref{sec:phase-space}. The `ID' column shows the original unique catalogue name of the stars; `\%' is the probability that the ten random realisations of the star hit the targeted phase-space area; and `type' indicates the type of the \textit{r}-PE stars.}
\end{table*}

\twocolumn


\end{appendix}

\end{document}